\documentclass[aps,prd,twocolumn,showpacs,amsmath,amssymb,superscriptaddress,altaffilletter,reprint]{revtex4-1}

\usepackage{graphicx,color}
\usepackage{lineno}
\usepackage{bm}
\usepackage{indentfirst}
\usepackage{fancyhdr}
\usepackage{subfigure}
\usepackage{amssymb}
\usepackage{amsmath}
\usepackage{multirow}
\usepackage{comment}
\usepackage{rotating}
\usepackage{enumerate}
\usepackage{dcolumn}
\usepackage{placeins}

\newcommand{\dmsq}{\Delta m^{2}}
\newcommand{\evsq}{\mathrm{eV}^{2}}
\newcommand{\gev}{\,\mathrm{GeV}}

\begin{document}
\pagewiselinenumbers

\title{\bf Dual baseline search for muon antineutrino disappearance at $0.1~\evsq < \dmsq < 100~\evsq$}

\affiliation{University of Alabama, Tuscaloosa, Alabama 35487, USA}
\affiliation{Argonne National Laboratory, Argonne, Illinois 60439, USA}
\affiliation{Institut de Fisica d'Altes Energies, Universitat Autonoma de Barcelona, E-08193 Bellaterra (Barcelona), Spain}
\affiliation{University of Cincinnati, Cincinnati, Ohio 45221, USA}
\affiliation{University of Colorado, Boulder, Colorado 80309, USA}
\affiliation{Columbia University, New York, New York 10027, USA}
\affiliation{Fermi National Accelerator Laboratory, Batavia, Illinois 60510, USA}
\affiliation{University of Florida, Gainesville, Florida 32611, USA}
\affiliation{High Energy Accelerator Research Organization (KEK), Tsukuba, Ibaraki 305-0801, Japan}
\affiliation{Imperial College London, London SW7 2AZ, United Kingdom}
\affiliation{Indiana University, Bloomington, Indiana 47405, USA}
\affiliation{Kamioka Observatory, Institute for Cosmic Ray Research, University of Tokyo, Gifu 506-1205, Japan}
\affiliation{Research Center for Cosmic Neutrinos, Institute for Cosmic Ray Research, University of Tokyo, Kashiwa, Chiba 277-8582, Japan}
\affiliation{Kyoto University, Kyoto 606-8502, Japan}
\affiliation{Los Alamos National Laboratory, Los Alamos, New Mexico 87545, USA}
\affiliation{Louisiana State University, Baton Rouge, Louisiana 70803, USA}
\affiliation{Massachusetts Institute of Technology, Cambridge, Massachusetts 02139, USA}
\affiliation{Instituto de Ciencias Nucleares, Universidad Nacional Aut\'onoma de M\'exico, D.F. 04510, M\'exico}
\affiliation{University of Michigan, Ann Arbor, Michigan 48109, USA}
\affiliation{Purdue University Calumet, Hammond, Indiana 46323, USA}
\affiliation{Universit$\grave{a}$ di Roma Sapienza, Dipartimento di Fisica and INFN, I-00185 Rome, Italy}
\affiliation{Saint Mary's University of Minnesota, Winona, Minnesota 55987, USA}
\affiliation{Tokyo Institute of Technology, Tokyo 152-8551, Japan}
\affiliation{Instituto de Fisica Corpuscular, Universidad de Valencia and CSIC, E-46071 Valencia, Spain}
\affiliation{Yale University, New Haven, Connecticut 06520, USA}

\author{G.~Cheng}\email[Corresponding author: ]{gcc2113@columbia.edu}\affiliation{Columbia University, New York, New York 10027, USA}
\author{W.~Huelsnitz}\email[Corresponding author: ]{whuelsn@fnal.gov}
\affiliation{Los Alamos National Laboratory, Los Alamos, New Mexico 87545, USA}
\author{A.~A. Aguilar-Arevalo}\affiliation{Instituto de Ciencias Nucleares, Universidad Nacional Aut\'onoma de M\'exico, D.F. 04510, M\'exico}
\author{J.~L.~Alcaraz-Aunion}\affiliation{Institut de Fisica d'Altes Energies, Universitat Autonoma de Barcelona, E-08193 Bellaterra (Barcelona), Spain}
\author{S.~J.~Brice}\affiliation{Fermi National Accelerator Laboratory, Batavia, Illinois 60510, USA}
\author{B.~C.~Brown}\affiliation{Fermi National Accelerator Laboratory, Batavia, Illinois 60510, USA}
\author{L.~Bugel}\affiliation{Massachusetts Institute of Technology, Cambridge, Massachusetts 02139, USA}
\author{J.~Catala-Perez}\affiliation{Instituto de Fisica Corpuscular, Universidad de Valencia and CSIC, E-46071 Valencia, Spain}
\author{E.~D.~Church}\affiliation{Yale University, New Haven, Connecticut 06520, USA}
\author{J.~M.~Conrad}\affiliation{Massachusetts Institute of Technology, Cambridge, Massachusetts 02139, USA}
\author{R.~Dharmapalan}\affiliation{University of Alabama, Tuscaloosa, Alabama 35487, USA}
\author{Z.~Djurcic}\affiliation{Argonne National Laboratory, Argonne, Illinois 60439, USA}
\author{U.~Dore}\affiliation{Universit$\grave{a}$ di Roma Sapienza, Dipartimento di Fisica and INFN, I-00185 Rome, Italy}
\author{D.~A.~Finley}\affiliation{Fermi National Accelerator Laboratory, Batavia, Illinois 60510, USA}
\author{R.~Ford}\affiliation{Fermi National Accelerator Laboratory, Batavia, Illinois 60510, USA}
\author{A.~J.~Franke}\affiliation{Columbia University, New York, New York 10027, USA}
\author{F.~G.~Garcia}\affiliation{Fermi National Accelerator Laboratory, Batavia, Illinois 60510, USA}
\author{G.~T.~Garvey}\affiliation{Los Alamos National Laboratory, Los Alamos, New Mexico 87545, USA}
\author{C.~Giganti}\altaffiliation[Present address: ]{DSM/Irfu/SPP, CEA Saclay, F-91191 Gif-sur-Yvette, France}\affiliation{Universit$\grave{a}$ di Roma Sapienza, Dipartimento di Fisica and INFN, I-00185 Rome, Italy}
\author{J.~J.~Gomez-Cadenas}\affiliation{Instituto de Fisica Corpuscular, Universidad de Valencia and CSIC, E-46071 Valencia, Spain}
\author{J.~Grange}\affiliation{University of Florida, Gainesville, Florida 32611, USA}
\author{P.~Guzowski}\altaffiliation[Present address: ]{The School of Physics and Astronomy, The University of Manchester, Manchester, M13 9PL, United Kingdom}\affiliation{Imperial College London, London SW7 2AZ, United Kingdom}
\author{A.~Hanson}\affiliation{Indiana University, Bloomington, Indiana 47405, USA}
\author{Y.~Hayato}\affiliation{Kamioka Observatory, Institute for Cosmic Ray Research, University of Tokyo, Gifu 506-1205, Japan}
\author{K.~Hiraide}\altaffiliation[Present address: ]{Kamioka Observatory, Institute for Cosmic Ray Research, University of Tokyo, Gifu 506-1205, Japan}\affiliation{Kyoto University, Kyoto 606-8502, Japan}
\author{C.~Ignarra}\affiliation{Massachusetts Institute of Technology, Cambridge, Massachusetts 02139, USA}
\author{R.~Imlay}\affiliation{Louisiana State University, Baton Rouge, Louisiana 70803, USA}
\author{R.~A. ~Johnson}\affiliation{University of Cincinnati, Cincinnati, Ohio 45221, USA}
\author{B.~J.~P.~Jones}\affiliation{Massachusetts Institute of Technology, Cambridge, Massachusetts 02139, USA}
\author{G.~Jover-Manas}\affiliation{Institut de Fisica d'Altes Energies, Universitat Autonoma de Barcelona, E-08193 Bellaterra (Barcelona), Spain}
\author{G.~Karagiorgi}\affiliation{Columbia University, New York, New York 10027, USA}\affiliation{Massachusetts Institute of Technology, Cambridge, Massachusetts 02139, USA}
\author{T.~Katori}\affiliation{Indiana University, Bloomington, Indiana 47405, USA}\affiliation{Massachusetts Institute of Technology, Cambridge, Massachusetts 02139, USA}
\author{Y.~K.~Kobayashi}\affiliation{Tokyo Institute of Technology, Tokyo 152-8551, Japan}
\author{T.~Kobilarcik}\affiliation{Fermi National Accelerator Laboratory, Batavia, Illinois 60510, USA}
\author{H.~Kubo}\affiliation{Kyoto University, Kyoto 606-8502, Japan}
\author{Y.~Kurimoto}\altaffiliation[Present address: ]{High Energy Accelerator Research Organization (KEK), Tsukuba, Ibaraki 305-0801, Japan}\affiliation{Kyoto University, Kyoto 606-8502, Japan}
\author{W.~C.~Louis}\affiliation{Los Alamos National Laboratory, Los Alamos, New Mexico 87545, USA}
\author{P.~F.~Loverre}\affiliation{Universit$\grave{a}$ di Roma Sapienza, Dipartimento di Fisica and INFN, I-00185 Rome, Italy}
\author{L.~Ludovici}\affiliation{Universit$\grave{a}$ di Roma Sapienza, Dipartimento di Fisica and INFN, I-00185 Rome, Italy}
\author{K.~B.~M.~Mahn}\altaffiliation[Present address: ]{TRIUMF, Vancouver, British Columbia, V6T 2A3, Canada}\affiliation{Columbia University, New York, New York 10027, USA}
\author{C.~Mariani}\altaffiliation[Present address: ]{Center for Neutrino Physics, Virginia Tech, Blacksburg, VA, USA}\affiliation{Columbia University, New York, New York 10027, USA}
\author{W.~Marsh}\affiliation{Fermi National Accelerator Laboratory, Batavia, Illinois 60510, USA}
\author{S.~Masuike}\affiliation{Tokyo Institute of Technology, Tokyo 152-8551, Japan}
\author{K.~Matsuoka}\affiliation{Kyoto University, Kyoto 606-8502, Japan}
\author{V.~T.~McGary}\affiliation{Massachusetts Institute of Technology, Cambridge, Massachusetts 02139, USA}
\author{W.~Metcalf}\affiliation{Louisiana State University, Baton Rouge, Louisiana 70803, USA}
\author{G.~B.~Mills}\affiliation{Los Alamos National Laboratory, Los Alamos, New Mexico 87545, USA}
\author{J.~Mirabal}\affiliation{Los Alamos National Laboratory, Los Alamos, New Mexico 87545, USA}
\author{G.~Mitsuka}\altaffiliation[Present address: ]{Solar-Terrestrial Environment Laboratory, Nagoya University, Furo-cho, Chikusa-ku, Nagoya, Japan}\affiliation{Research Center for Cosmic Neutrinos, Institute for Cosmic Ray Research, University of Tokyo, Kashiwa, Chiba 277-8582, Japan}
\author{Y.~Miyachi}\altaffiliation[Present address: ] {Yamagata University, Yamagata, 990-8560 Japan}\affiliation{Tokyo Institute of Technology, Tokyo 152-8551, Japan}
\author{S.~Mizugashira}\affiliation{Tokyo Institute of Technology, Tokyo 152-8551, Japan}
\author{C.~D.~Moore}\affiliation{Fermi National Accelerator Laboratory, Batavia, Illinois 60510, USA}
\author{J.~Mousseau}\affiliation{University of Florida, Gainesville, Florida 32611, USA}
\author{Y.~Nakajima}\altaffiliation[Present address: ]{Lawrence Berkeley National Laboratory, Berkeley, CA 94720, USA}\affiliation{Kyoto University, Kyoto 606-8502, Japan}
\author{T.~Nakaya}\affiliation{Kyoto University, Kyoto 606-8502, Japan}
\author{R.~Napora}\altaffiliation[Present address: ]{Epic Systems, Inc.}\affiliation{Purdue University Calumet, Hammond, Indiana 46323, USA}
\author{P.~Nienaber}\affiliation{Saint Mary's University of Minnesota, Winona, Minnesota 55987, USA}
\author{D.~Orme}\affiliation{Kyoto University, Kyoto 606-8502, Japan}
\author{B.~Osmanov}\affiliation{University of Florida, Gainesville, Florida 32611, USA}
\author{M.~Otani}\affiliation{Kyoto University, Kyoto 606-8502, Japan}
\author{Z.~Pavlovic}\affiliation{Los Alamos National Laboratory, Los Alamos, New Mexico 87545, USA}
\author{D.~Perevalov}\affiliation{Fermi National Accelerator Laboratory, Batavia, Illinois 60510, USA}
\author{C.~C.~Polly}\affiliation{Fermi National Accelerator Laboratory, Batavia, Illinois 60510, USA}
\author{H.~Ray}\affiliation{University of Florida, Gainesville, Florida 32611, USA}
\author{B.~P.~Roe}\affiliation{University of Michigan, Ann Arbor, Michigan 48109, USA}
\author{A.~D.~Russell}\affiliation{Fermi National Accelerator Laboratory, Batavia, Illinois 60510, USA}
\author{F.~Sanchez}\affiliation{Institut de Fisica d'Altes Energies, Universitat Autonoma de Barcelona, E-08193 Bellaterra (Barcelona), Spain}
\author{M.~H.~Shaevitz}\affiliation{Columbia University, New York, New York 10027, USA}
\author{T.-A.~Shibata}\affiliation{Tokyo Institute of Technology, Tokyo 152-8551, Japan}
\author{M.~Sorel}\affiliation{Instituto de Fisica Corpuscular, Universidad de Valencia and CSIC, E-46071 Valencia, Spain}
\author{J.~Spitz}\affiliation{Massachusetts Institute of Technology, Cambridge, Massachusetts 02139, USA}
\author{I.~Stancu}\affiliation{University of Alabama, Tuscaloosa, Alabama 35487, USA}
\author{R.~J.~Stefanski}\affiliation{Fermi National Accelerator Laboratory, Batavia, Illinois 60510, USA}
\author{H.~Takei}\altaffiliation[Present address: ]{Kitasato University, Tokyo, 108-8641 Japan}\affiliation{Tokyo Institute of Technology, Tokyo 152-8551, Japan}
\author{H.-K.~Tanaka}\affiliation{Massachusetts Institute of Technology, Cambridge, Massachusetts 02139, USA}
\author{M.~Tanaka}\affiliation{High Energy Accelerator Research Organization (KEK), Tsukuba, Ibaraki 305-0801, Japan}
\author{R.~Tayloe}\affiliation{Indiana University, Bloomington, Indiana 47405, USA}
\author{I.~J.~Taylor}\altaffiliation[Present address: ]{Department of Physics and Astronomy, State University of New York, Stony Brook, New York 11794-3800, USA }\affiliation{Imperial College London, London SW7 2AZ, United Kingdom}
\author{R.~J.~Tesarek}\affiliation{Fermi National Accelerator Laboratory, Batavia, Illinois 60510, USA}
\author{Y.~Uchida}\affiliation{Imperial College London, London SW7 2AZ, United Kingdom}
\author{R.~G.~Van~de~Water}\affiliation{Los Alamos National Laboratory, Los Alamos, New Mexico 87545, USA}
\author{J.~J.~Walding}\altaffiliation[Present address: ]{Department of Physics, Royal Holloway, University of London, Egham, TW20 0EX, United Kingdom}\affiliation{Imperial College London, London SW7 2AZ, United Kingdom}
\author{M.~O.~Wascko}\affiliation{Imperial College London, London SW7 2AZ, United Kingdom}
\author{D.~H.~White}\affiliation{Los Alamos National Laboratory, Los Alamos, New Mexico 87545, USA}
\author{H.~B.~White}\affiliation{Fermi National Accelerator Laboratory, Batavia, Illinois 60510, USA}
\author{D.~A.~Wickremasinghe}\affiliation{University of Cincinnati, Cincinnati, Ohio 45221, USA}
\author{M.~Yokoyama}\altaffiliation[Present address: ]{Department of Physics, University of Tokyo, Tokyo 113-0033, Japan}\affiliation{Kyoto University, Kyoto 606-8502, Japan}
\author{G.~P.~Zeller}\affiliation{Fermi National Accelerator Laboratory, Batavia, Illinois 60510, USA}
\author{E.~D.~Zimmerman}\affiliation{University of Colorado, Boulder, Colorado 80309, USA}

\collaboration{MiniBooNE and SciBooNE Collaborations}\noaffiliation

\begin{abstract}
The MiniBooNE and SciBooNE collaborations report the results of a joint search for short baseline disappearance of $\bar{\nu}_{\mu}$ at Fermilab's Booster Neutrino Beamline. The MiniBooNE Cherenkov detector and the SciBooNE tracking detector observe antineutrinos from the same beam, therefore the combined analysis of their datasets serves to partially constrain some of the flux and cross section uncertainties.  Uncertainties in the $\nu_{\mu}$ background were constrained by neutrino flux and cross section measurements performed in both detectors.  A likelihood ratio method was used to set a $90\%$ confidence level upper limit on $\bar{\nu}_{\mu}$ disappearance that dramatically improves upon prior limits in the $\dmsq$=0.1--100 $\evsq$ region.
\end{abstract}

\date{\today}

\pacs{14.60.Lm, 14.60.Pq, 14.60.St}

\maketitle

\section{Introduction}

Recently there has been increasing evidence in support of neutrino oscillations in the $\Delta m^{2}$ $\approx 1$ $\mathrm{eV}^{2}$ region.  The LSND~\cite{PhysRevD.64.112007} experiment observed an excess of $\bar\nu_{e}$-like events in a $\bar{\nu}_{\mu}$ beam. MiniBooNE~\cite{AguilarArevalo:2007it,AguilarArevalo:2008rc,PhysRevLett.105.181801} has observed an excess of $\nu_e$-like and $\bar{\nu}_e$-like events, in a $\nu_\mu$ beam and $\bar{\nu}_{\mu}$ beam, respectively.  Additional evidence for short-baseline anomalies with $L/E \approx1$, where $L$ is the neutrino path length in km and $E$ the neutrino energy in GeV, includes the deficit of events observed in reactor antineutrino experiments~\cite{Mention:2011rk} and radioactive source neutrino measurements~\cite{PhysRevC.83.065504}. If these anomalies are due to neutrino oscillations in the $\Delta m^{2}$ $\approx 1$ $\mathrm{eV}^{2}$ range, then they could imply the existence of one or more new sterile neutrino species that do not participate in standard weak interactions but mix with the known neutrino flavors through additional mass eigenstates.  Observation of $\nu_\mu$ $\left( \bar{\nu}_{\mu}\right)$ disappearance in conjunction with $\nu_e$ $\left(\bar{\nu}_e \right)$ appearance in this $\Delta m^{2}$ range would be a smoking-gun for the presence of these sterile neutrinos.  Alternatively, constraining $\nu_\mu$ $\left( \bar{\nu}_{\mu}\right)$ disappearance can, along with global $\nu_e$ $\left(\bar{\nu}_e \right)$ disappearance data, constrain the oscillation interpretation of the $\nu_e$ $\left(\bar{\nu}_e \right)$ appearance signals in LSND and MiniBooNE~\cite{MB_appearance}.

Searches for $\nu_{\mu}$ and $\bar{\nu}_{\mu}$ disappearance in MiniBooNE were performed in 2009~\cite{PhysRevLett.103.061802}. No evidence for disappearance was found. The search for $\nu_{\mu}$ disappearance was recently repeated in MiniBooNE with the inclusion of data from the SciBooNE detector in a joint analysis~\cite{Mahn:2011ea}.  Once again, the results were consistent with no $\nu_{\mu}$ disappearance. The analysis presented here is an improved search for $\bar{\nu}_{\mu}$ disappearance using data from MiniBooNE and SciBooNE taken while the Booster Neutrino Beamline (BNB) operated in antineutrino mode.

The Monte Carlo (MC) predictions for both MiniBooNE and SciBooNE were updated to account for recent neutrino flux and cross section measurements made with both experiments.  The data from both detectors were then simultaneously fit to a simple two-antineutrino oscillation model. Improved constraints on MC predictions, the inclusion of SciBooNE data, and a MiniBooNE antineutrino data set nearly 3 times larger than what was available for the original $\bar\nu_{\mu}$ disappearance analysis, have allowed a $90\%$ confidence level upper limit to be set that dramatically improves upon prior limits in the $\dmsq$ = 0.1--100 $\evsq$ region, pushing down into the region of parameter space of interest to sterile neutrino models.

This paper is organized as follows. Section~\ref{sec:beamlinedetectors} describes the BNB and the MiniBooNE and SciBooNE detectors. Then, the simulation of neutrino interactions with nuclei and subsequent detector responses are described in Sec.~\ref{sec:mcsim}. The event selection and reconstruction for both detectors are described in Sec.~\ref{sec:eventselectionreconstruction}. The parameters for the MC tuning and its systematic uncertainties are given in Sec.~\ref{sec:sysuncertain}. Section~\ref{sec:method} describes the analysis methodology. The results of the analysis are presented in Sec.~\ref{sec:results}, and the
final conclusions are given in Sec.~\ref{sec:conclusions}.

\section{Beamline and Experimental Apparatus}\label{sec:beamlinedetectors}

\begin{figure*}[htbn!]
\begin{center}
\includegraphics[width=2.0\columnwidth]{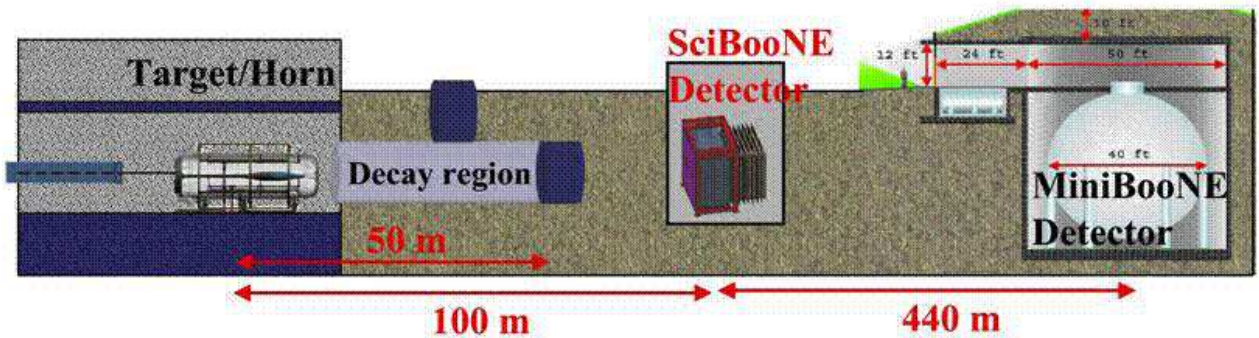}
\caption{Schematic view of the BNB from the beryllium target and magnetic horn to the SciBooNE and MiniBooNE detectors.}
\label{fig:detector_setup}
\end{center}
\end{figure*}

MiniBooNE and SciBooNE both use the BNB at Fermilab in Batavia, Illinois. The 8 GeV kinetic energy protons from the booster accelerator strike a 1.7 interaction length beryllium target, which is located inside a focusing horn. The horn is pulsed in time with the beam to produce a toroidal magnetic field that, depending on the polarity setting, will either focus $\pi^-$/$K^-$ and defocus $\pi^+$/$K^+$ or vice-versa. These mesons then pass through a 60 cm long collimator and decay in flight along a 50 m long tunnel. A schematic view of the BNB from the beryllium target to both detectors is shown in Fig.~\ref{fig:detector_setup}.

\begin{figure*}[htbn!]
\begin{center}
\subfigure[~MiniBooNE antineutrino mode flux]{\includegraphics[width=1.0\columnwidth]{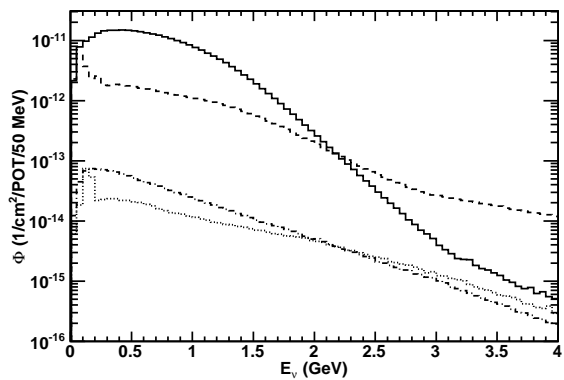}}
\subfigure[~SciBooNE antineutrino mode flux]{\includegraphics[width=1.0\columnwidth]{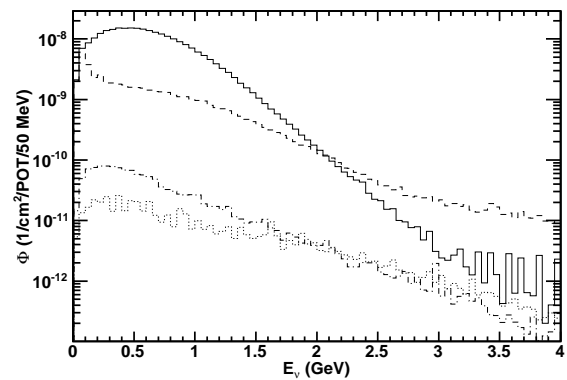}}
\caption{The neutrino and antineutrino flux prediction as a function of true neutrino(antineutrino) energy, in antineutrino mode at the MiniBooNE and SciBooNE detectors. The $\bar\nu_{\mu}$ flux is represented by the solid line, the $\nu_{\mu}$ flux is represented by the dashed line, the $\bar\nu_{e}$ flux is represented by the dot-dashed line, and the $\nu_{e}$ flux is represented by the dotted line.}
\label{fig:flux_antinu}
\end{center}
\end{figure*}

The resulting neutrino beam will have an enhanced flux of either muon neutrinos (\textit{neutrino mode}) or muon antineutrinos (\textit{antineutrino mode}). In antineutrino mode beam running, the flux of antineutrinos in the beam will be referred to as the right-sign (RS) flux and the flux of neutrinos in the beam will be referred to as the wrong-sign (WS) flux. These two designations are used because antineutrinos are the signal in this analysis and neutrinos are an intrinsic background. Figure~\ref{fig:flux_antinu} shows the neutrino and antineutrino flux prediction in antineutrino mode at both the MiniBooNE and SciBooNE detectors. Details on the beamline and flux predictions are given in Ref.~\cite{AguilarArevalo:2008yp}.

\begin{figure}[htbn!]
\begin{center}
\includegraphics[width=0.7\columnwidth]{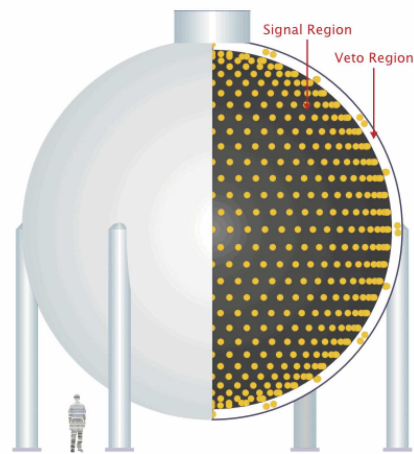}
\caption{Schematic view of the MiniBooNE detector.}
\label{fig:mb_detector}
\end{center}
\end{figure}

The MiniBooNE detector~\cite{AguilarArevalo:2008qa} is located 541 m downstream of the antineutrino production target and consists of a spherical 12.2 m diameter tank containing 800 tons of mineral oil ($\rm CH_2$), beneath at least 3 m of earth overburden. The fiducial volume is a sphere 10 m in diameter, with a fiducial mass of 450 tons.  The detector is instrumented with 1280 8 inch photomultiplier tubes (PMTs) in the active region, and 240 8 inch PMTs in an outer, veto region.  Events are reconstructed based on timing and charge information mostly from Cherenkov radiation. A schematic of the MiniBooNE detector is shown in Fig.~\ref{fig:mb_detector}.

\begin{figure}[htbn!]
\begin{center}
\includegraphics[width=0.9\columnwidth]{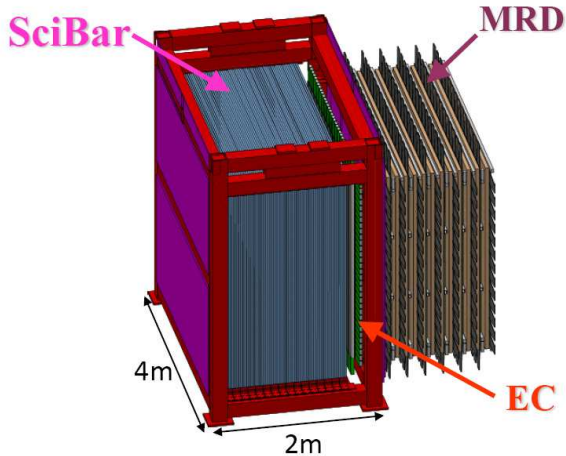}
\caption{Schematic view of the SciBooNE detector.}
\label{fig:sb_detector}
\end{center}
\end{figure}

The SciBooNE detector~\cite{Hiraide:2008eu} is located 100 m downstream of the target. SciBooNE is a discrete tracking detector comprised of three subdetectors (in order from upstream to downstream): a fully active and finely segmented scintillator tracker (SciBar), an electromagnetic calorimeter (EC), and a muon range detector (MRD). The SciBar subdetector~\cite{Nitta:2004nt} consists of 14336 extruded polystyrene ($\rm C_8H_8$) strips arranged vertically and horizontally to construct a $\rm 3 \times 3 \times 1.7 m^3$ volume. Each scintillator strip is read out by a wavelength shifting fiber attached to a 64-channel multianode PMT (MA-PMT). The 15 ton SciBar subdetector (10.6 ton fiducial volume) provides the primary interaction target. The EC subdetector is a two plane (vertical and horizontal) ``spaghetti''-type calorimeter; 64 modules made of 1~mm scintillating fibers embedded in lead foil are bundled and read out at both ends by PMTs. The MRD subdetector, designed to measure muon momentum, is made from 12 iron plates, each 5 cm thick, sandwiched between 13 alternating horizontal and vertical scintillator planes of thickness 6 mm that are read out via 362 individual 2 inch PMTs. A schematic of the SciBooNE detector is shown in Fig.~\ref{fig:sb_detector}.

\section{Monte Carlo Simulation}\label{sec:mcsim}

Simulation of the neutrino and antineutrino flux, neutrino and antineutrino interactions in the detector, and detector response has been discussed in detail in previous publications for MiniBooNE~\cite{PhysRevD.81.092005,PhysRevD.84.072005} and SciBooNE~\cite{Nakajima:2010fp}.  Calculation of the neutrino and antineutrino flux at the detector is done with a GEANT4-based model~\cite{Agostinelli:2002hh} that is constrained by external measurements~\cite{Catanesi:2007ab,AguilarArevalo:2008yp} and accounts for proton transport to the target, p-Be interactions in the target, meson production, focusing by the magnetic horn, meson propagation and decay, and neutrino and antineutrino propagation to the detectors.

Neutrino and antineutrino interactions in both detectors are simulated using the NUANCE~\cite{Casper:2002sd} event generator.  Bound nucleons are described by the relativistic Fermi gas (RFG) model~\cite{Smith:1972xh}. The MiniBooNE detector response is simulated using GEANT3~\cite{Brun:1994aa}, which takes the final-state particles emerging from a nucleus and propagates them through the detector.  The GEANT3 code was modified to include a custom model for light propagation in the detector~\cite{BrownIEEE:2004} and to use GCALOR~\cite{Zeitnitz:1994bs} for pion absorption and charge exchange in the detector medium.  SciBooNE uses GEANT4~\cite{Heikkinen:2003sc} to simulate the interactions of hadronic particles with detector materials.

\section{Event Selection and Reconstruction}\label{sec:eventselectionreconstruction}

MiniBooNE data from a total of $1.01 \times 10^{21}$ protons on target (POT) operation in antineutrino mode, from July 2006 up through April 2012, are included in the analysis. Data from SciBooNE antineutrino mode operation from June 2007 through August 2008 are included, comprising a total of $1.53 \times 10^{20}$ POT for the SciBooNE contribution.

MiniBooNE event selection and reconstruction is essentially identical to that used for a previous neutrino mode $\nu_\mu$ cross section measurement~\cite{PhysRevD.81.092005}.  Events with only a single $\mu^{+}$ in the detector are selected.  Event selection cuts are based on the beam timing, fiducial volume, observation of two correlated events (the muon and its decay electron), and the likelihood of the fit to the muon hypothesis. These cuts are designed to reject incoming particles (i.e. muons from cosmic rays or from neutrino and antineutrino interactions in the surrounding material), ensure that the event is contained within the detector, and ensure correct event classification as well as accurate muon energy estimation. The capture of $\mu^{-}$ resulting from initial $\nu_\mu$ charged current quasielastic (CCQE) interaction events is simulated in the MC and these specific events are not selected. In antineutrino mode, a sizable fraction of the events (roughly $20\%$) are due to $\nu_\mu$ interactions.  MiniBooNE cannot distinguish between $\nu_\mu$ and $\bar\nu_{\mu}$ events on an event-by-event basis, so $\mu^-$s from $\nu_\mu$ interactions are an irreducible background.

For SciBooNE, the event selection and reconstruction is nearly identical to the previous inclusive charged current measurement~\cite{Nakajima:2010fp}. Two-dimensional SciBar tracks are reconstructed using a cellular automaton algorithm~\cite{Aunion:2010zz} from SciBar hits. Three-dimensional SciBar tracks are then reconstructed based on the timing and end point positions of the two-dimensional SciBar tracks. Two-dimensional tracks in the MRD are independently reconstructed using hits in the MRD that are clustered within a 50 ns timing window. Three-dimensional tracks in the MRD are reconstructed by matching the timing of the two-dimensional projections.  If the downstream edge of a SciBar track lies in the last two layers of SciBar, a search for a matching track or hits in the MRD is performed.  The upstream edge of the MRD track is required to be on either one of the first two layers of the MRD, and to be within 30 cm of the projected entry point of the SciBar track into the MRD (a more detailed description of the track reconstruction can be found in Ref.~\cite{Hiraide:2008eu}).

To select $\mu^{+}$ events, the highest momentum track per event in the beam on-time window is required to have $p_\mu>0.25~\gev/c$ to reduce the number of neutral current (NC) events. The energy loss of the track in SciBar must be consistent with a muon hypothesis, and must originate within the 10.6~ton SciBar fiducial volume. These muon candidate tracks are further categorized as SciBar-stopped or MRD-stopped. SciBar-stopped events have the downstream end point of the muon candidate track contained in the SciBar fiducial volume. MRD-stopped events have the muon candidate track being a SciBar track matched to MRD hits or to an MRD track with a downstream end point that does not exit the back or sides of the MRD. Both SciBar-stopped and MRD-stopped events are used in the analysis. SciBooNE has no overburden so cosmic backgrounds must be subtracted. For cosmic background estimation, the same muon selection criteria are applied to a beam-off time window that is 5 times longer than the beam-on window. This event rate is scaled and subtracted from the beam-on data.

The selected events include $\bar\nu_{\mu}$ and $\nu_\mu$ interactions on carbon and hydrogen in the detectors. The reconstructed antineutrino energy is based on the assumption that the interaction is always a $\bar\nu_{\mu}$ CCQE interaction with a proton at rest in carbon: $\bar\nu_{\mu} + p \rightarrow \mu + n$. Hence, it is a function of the measured energy and direction of the outgoing muon.  The equation for reconstructed energy is
\begin{equation}
E_\nu ^{QE} = \frac{{M_n^2  - \left( {M_p  - E_B} \right)^2  - M_\mu ^2  + 2\left( {M_p  - E_B} \right)E_\mu  }}{{2\left( {M_p  - E_B - E_\mu   + P_\mu  \cos \theta _\mu  } \right)}},
\label{eq:reco_energy}
\end{equation}
where $M_{n}$ and $M_{p}$ are the mass of the neutron and proton, $M_{\mu}$, $E_{\mu}$, $P_{\mu}$, and $\theta_{\mu}$ are the mass, energy, momentum, and direction of the outgoing muon, and $E_B$ is the binding energy (30 MeV for protons in carbon). Equation~\ref{eq:reco_energy} is applied to all selected events in data and MC, even though a sizable fraction of the events are not CCQE [i.e. charged current single $\pi$ (CC1$\pi$), charged current multipion (CC multipion), or NC events misidentified as CCQE events]. The impact of the CCQE reconstruction assumption, which leads to reduced accuracy in reconstructed energy for non-CCQE events, is accounted for in MC, which also includes these selected non-CCQE events. MiniBooNE has an estimated resolution for reconstructed energy of $8.3\%$ for CCQE events and $13.9\%$ for all events. SciBooNE has an estimated reconstructed energy resolution of $9.6\%$ for CCQE events and $24.6\%$ for all events.

MiniBooNE and SciBooNE data and MC are put in 21-bin histograms of $E_{\nu}^{QE}$.  The binning goes from 300 MeV to 1.9 GeV, with individual bin widths as follows: bin 1, 100 MeV; bins 2-19, 66.7 MeV; bin 20, 100 MeV; bin 21, 200 MeV.  The first and last two bins are wider to ensure adequate event statistics in data and MC.

\begin{figure*}[htbn!]
\begin{center}
\subfigure[~MiniBooNE RS Events]{\includegraphics[width=0.99\columnwidth]{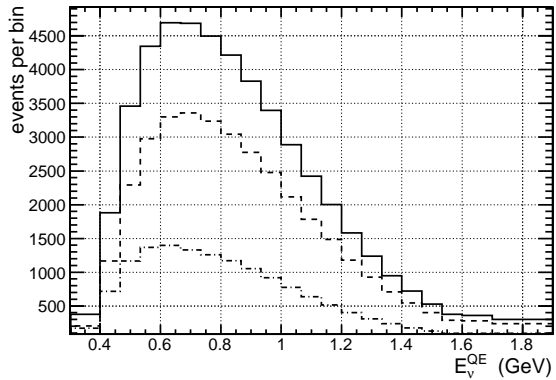}}
\subfigure[~MiniBooNE WS Events]{\includegraphics[width=0.99\columnwidth]{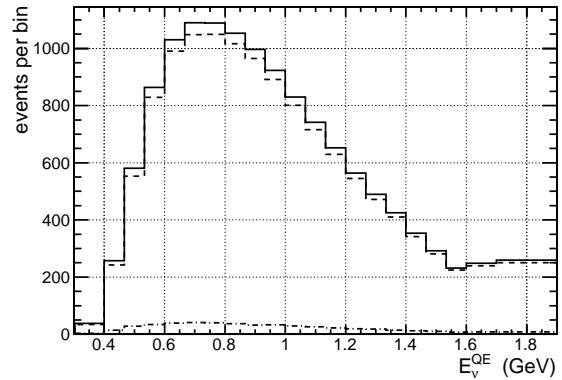}}
\caption{Reconstructed antineutrino and neutrino energy ($E_{\nu}^{QE}$) distributions for selected RS and WS MiniBooNE events on different target types (hydrogen or carbon) from MiniBooNE MC. Total events are represented by the solid line, events with interaction on carbon are represented by the dashed line, and events with interaction on hydrogen are represented by the dot-dashed line.}
\label{fig:mb_target}
\end{center}
\end{figure*}

\begin{figure*}[htbn!]
\begin{center}
\subfigure[~MiniBooNE RS Events]{\includegraphics[width=0.99\columnwidth]{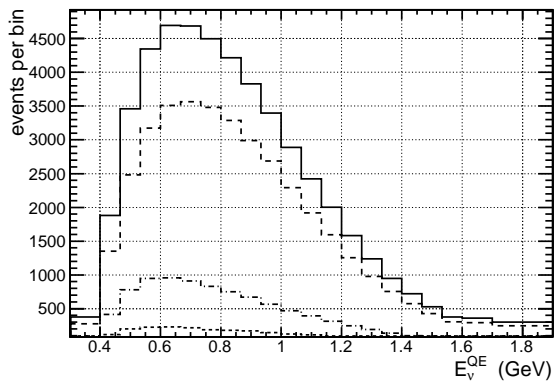}}
\subfigure[~MiniBooNE WS Events]{\includegraphics[width=0.99\columnwidth]{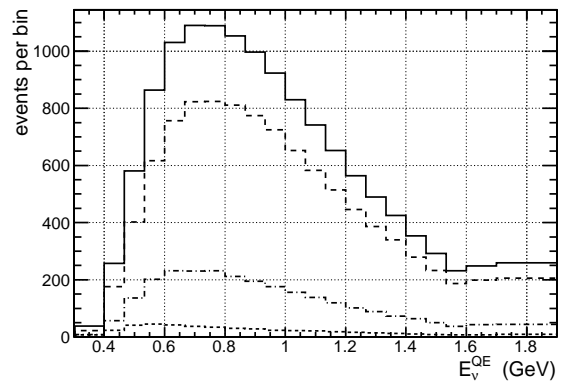}}
\caption{Reconstructed antineutrino and neutrino energy ($E_{\nu}^{QE}$) distributions for selected RS and WS MiniBooNE events for different interaction types (CCQE, CC$1\pi$, other) from MiniBooNE MC.  Total events are represented by the solid line, CCQE interaction events are represented by the dashed line, CC$1\pi$ interaction events are represented by the dot-dashed line, and all other interaction (CC multi-$\pi$ or NC) events are represented by the short-dashed line.}
\label{fig:mb_type}
\end{center}
\end{figure*}

Figure~\ref{fig:mb_target} shows the predicted event distributions in MiniBooNE for reconstructed antineutrino and neutrino energy, for events on hydrogen and carbon nuclei.  Figure~\ref{fig:mb_type} shows the predictions for MiniBooNE's reconstructed antineutrino and neutrino energy distributions by interaction type: CCQE, CC$1\pi$, and all other interaction types (CC multipion and NC). Table~\ref{tab:sel_event_type} shows the MC predictions for the selected MiniBooNE events by neutrino and interaction type.  $\nu_e$ and $\bar\nu_{e}$ contamination is negligible.

\begin{figure*}[htbn!]
\begin{center}
\subfigure[~SciBooNE RS Events]{\includegraphics[width=0.99\columnwidth]{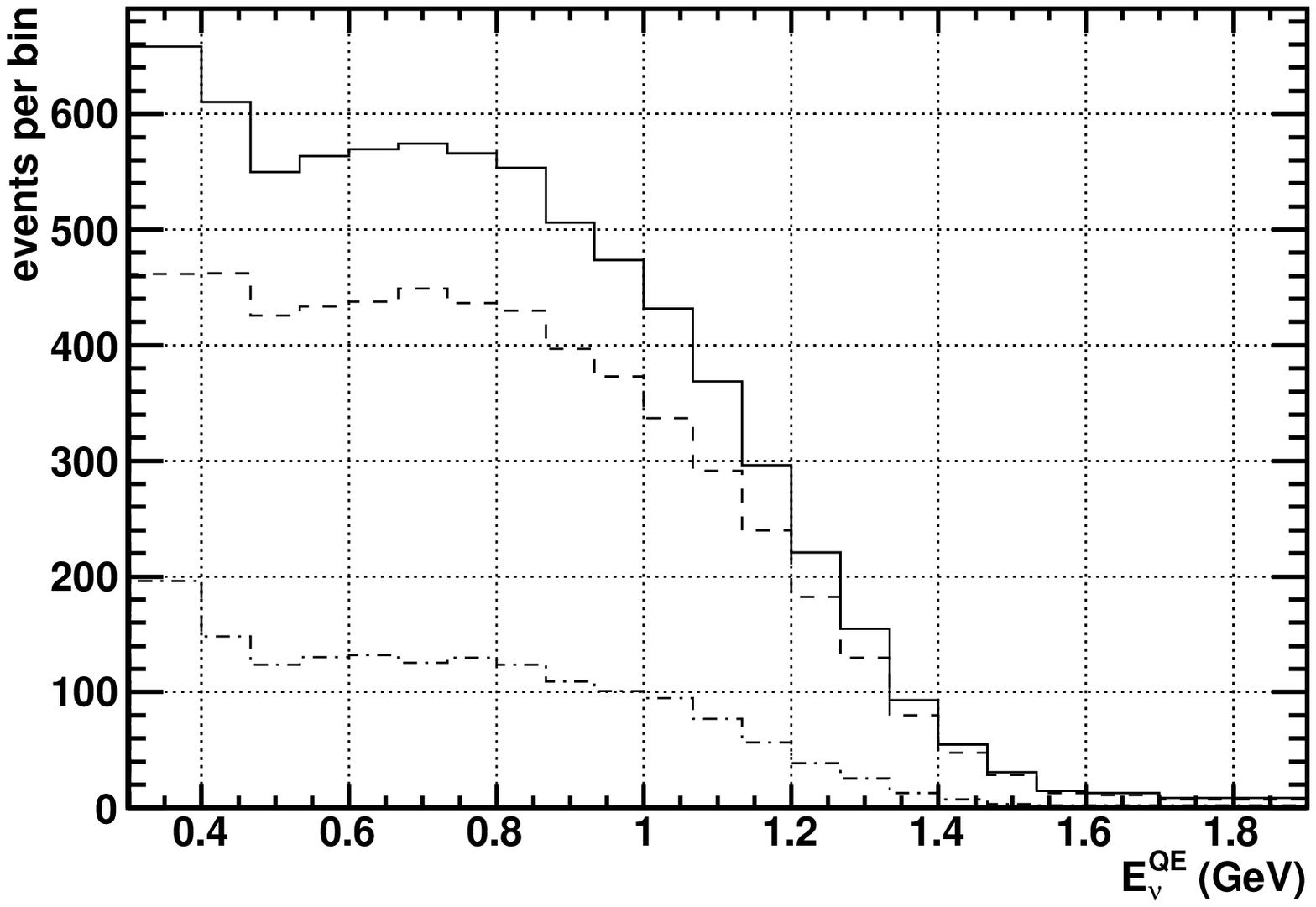}}
\subfigure[~SciBooNE WS Events]{\includegraphics[width=0.99\columnwidth]{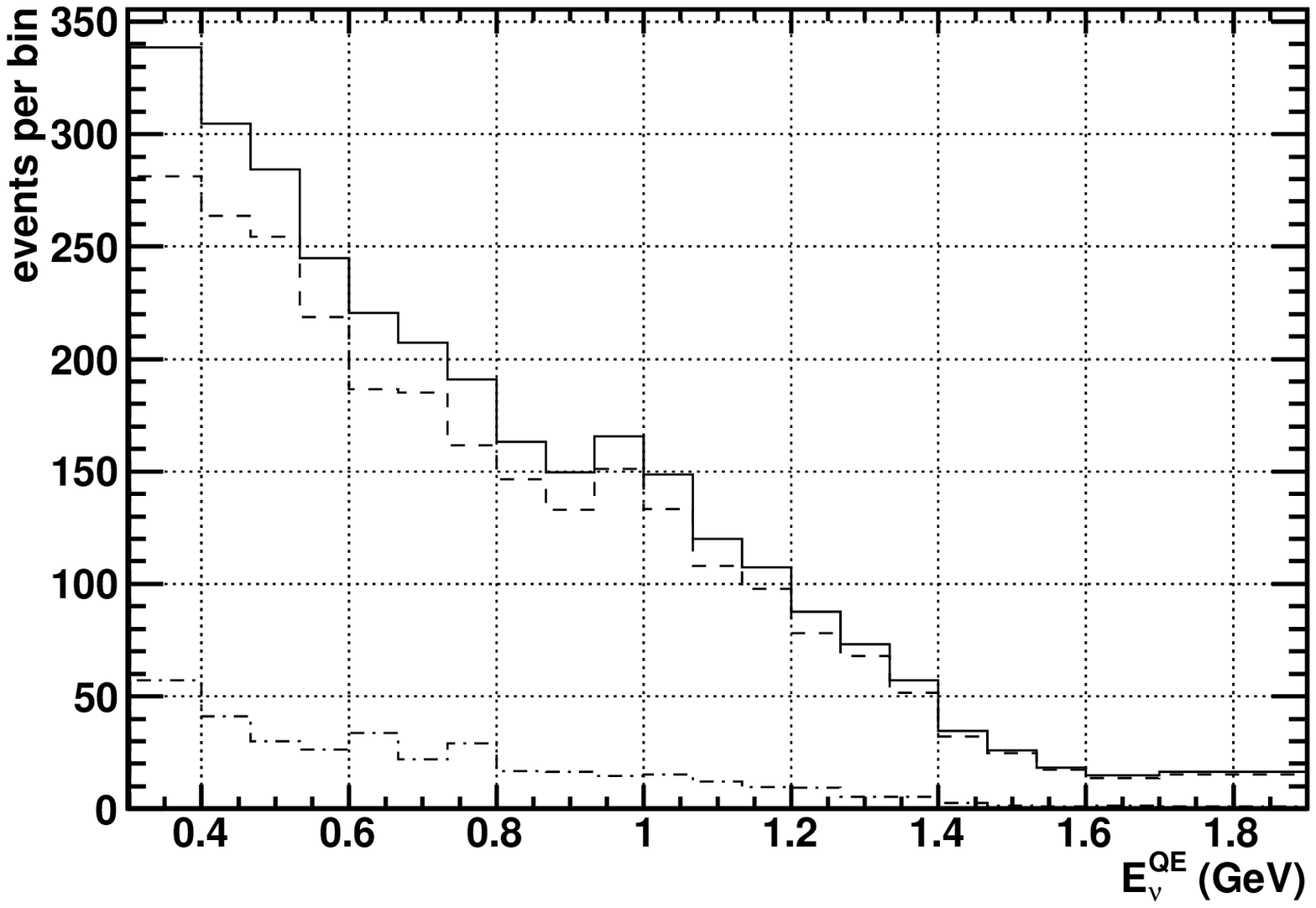}}
\caption{Reconstructed antineutrino and neutrino energy ($E_{\nu}^{QE}$) distributions for selected RS and WS SciBooNE events on different target types (hydrogen or carbon) from SciBooNE MC. Total events are represented by the solid line, events with interaction on carbon are represented by the dashed line, and events with interaction on hydrogen are represented by the dot-dashed line.}
\label{fig:sb_target}
\end{center}
\end{figure*}

\begin{figure*}[htbn!]
\begin{center}
\subfigure[~SciBooNE RS Events]{\includegraphics[width=0.99\columnwidth]{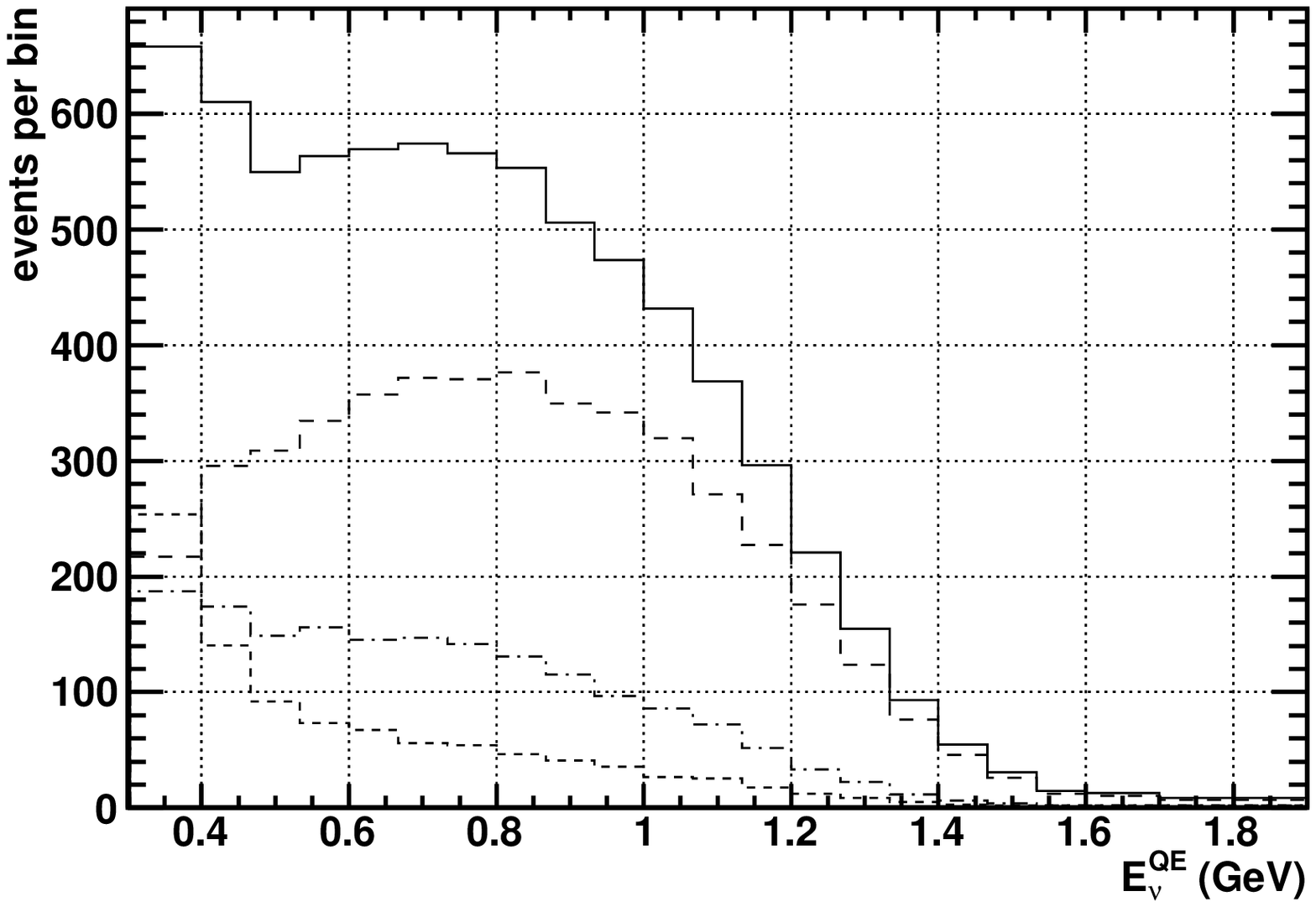}}
\subfigure[~SciBooNE WS Events]{\includegraphics[width=0.99\columnwidth]{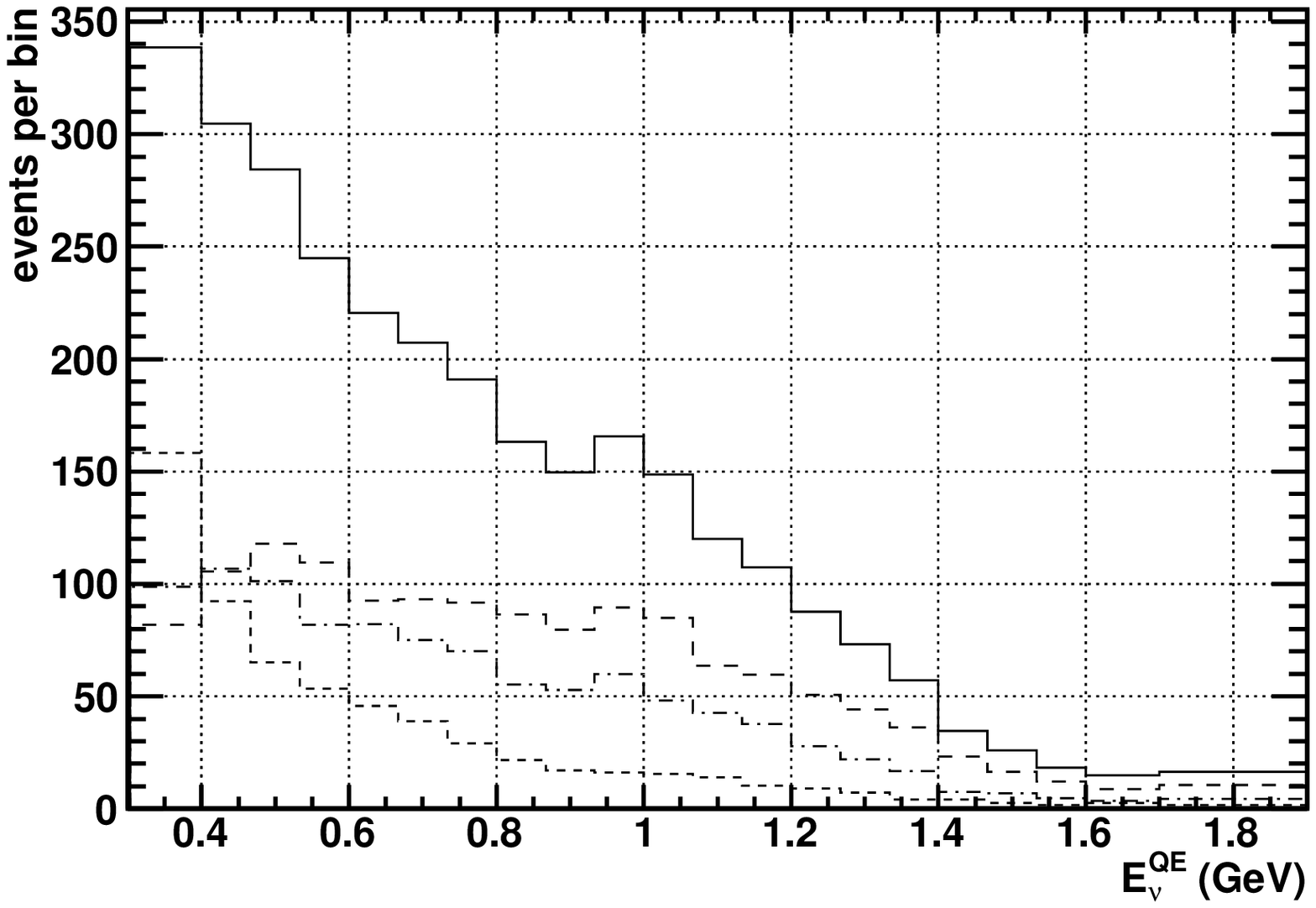}}
\caption{Reconstructed antineutrino and neutrino energy ($E_{\nu}^{QE}$) distributions for selected RS and WS SciBooNE events for different interaction types (CCQE, CC$1\pi$, other) from SciBooNE MC.  Total events are represented by the solid line, CCQE interaction events are represented by the dashed line, CC$1\pi$ interaction events are represented by the dot-dashed line, and all other interaction (CC multi-$\pi$ or NC) events are represented by the short-dashed line.}
\label{fig:sb_type}
\end{center}
\end{figure*}

The following plots show several properties of the selected SciBooNE events, as predicted by simulation.  Figure~\ref{fig:sb_target} shows the reconstructed antineutrino and neutrino energy distributions for events on hydrogen and carbon nuclei. Figure~\ref{fig:sb_type} shows the reconstructed antineutrino and neutrino energy distribution by interaction type: CCQE, CC$1\pi$, and other (CC multipion and NC). Table~\ref{tab:sel_event_type} shows the MC predictions for the selected SciBooNE events by neutrino and interaction type. The data set is estimated to contain an additional 811 events from cosmic ray muons.

\begin{table}[htbn!]
\caption{MC predictions for the number of selected events by neutrino and interaction type in both MiniBooNE and SciBooNE.}
\begin{tabular}{l|c|c|c|c}
\hline\hline
                      & \multicolumn{2}{c|}{MiniBooNE} & \multicolumn{2}{c}{SciBooNE}  \\\hline
interaction type      & $\bar\nu$ events   & $\nu$ events        & $\bar\nu$ events  & $\nu$ events   \\ \hline
CCQE                  & 37428       & 9955         & 4619       & 1359    \\ \hline
CC$1\pi$              & 8961        & 2593         & 1735       & 1006    \\ \hline
CC multi-$\pi$ or NC  & 2364        & 460          & 959        & 610     \\
\hline\hline
\end{tabular}
\label{tab:sel_event_type}
\end{table}

The difference in shape between the SciBooNE RS (WS) and MiniBooNE RS (WS) energy distributions is mainly due to different event selection criteria between MiniBooNE and SciBooNE. MiniBooNE selects for CCQE interaction events and SciBooNE selects for all CC interaction events so the SciBooNE sample has a larger percentage of non-CCQE interaction events. Since the antineutrino energy reconstruction is based on a CCQE interaction assumption, there are more SciBooNE events with a larger discrepancy between true antineutrino energy and reconstructed antineutrino energy than in MiniBooNE, leading to shape differences. Differences in selection efficiency, antineutrino flux at the detector locations, and background rejection between MiniBooNE and SciBooNE also contribute to the shape differences.

\begin{figure}[htbn!]
\begin{center}
\subfigure[~MiniBooNE (anti)neutrino path lengths]{\includegraphics[width=0.95\columnwidth]{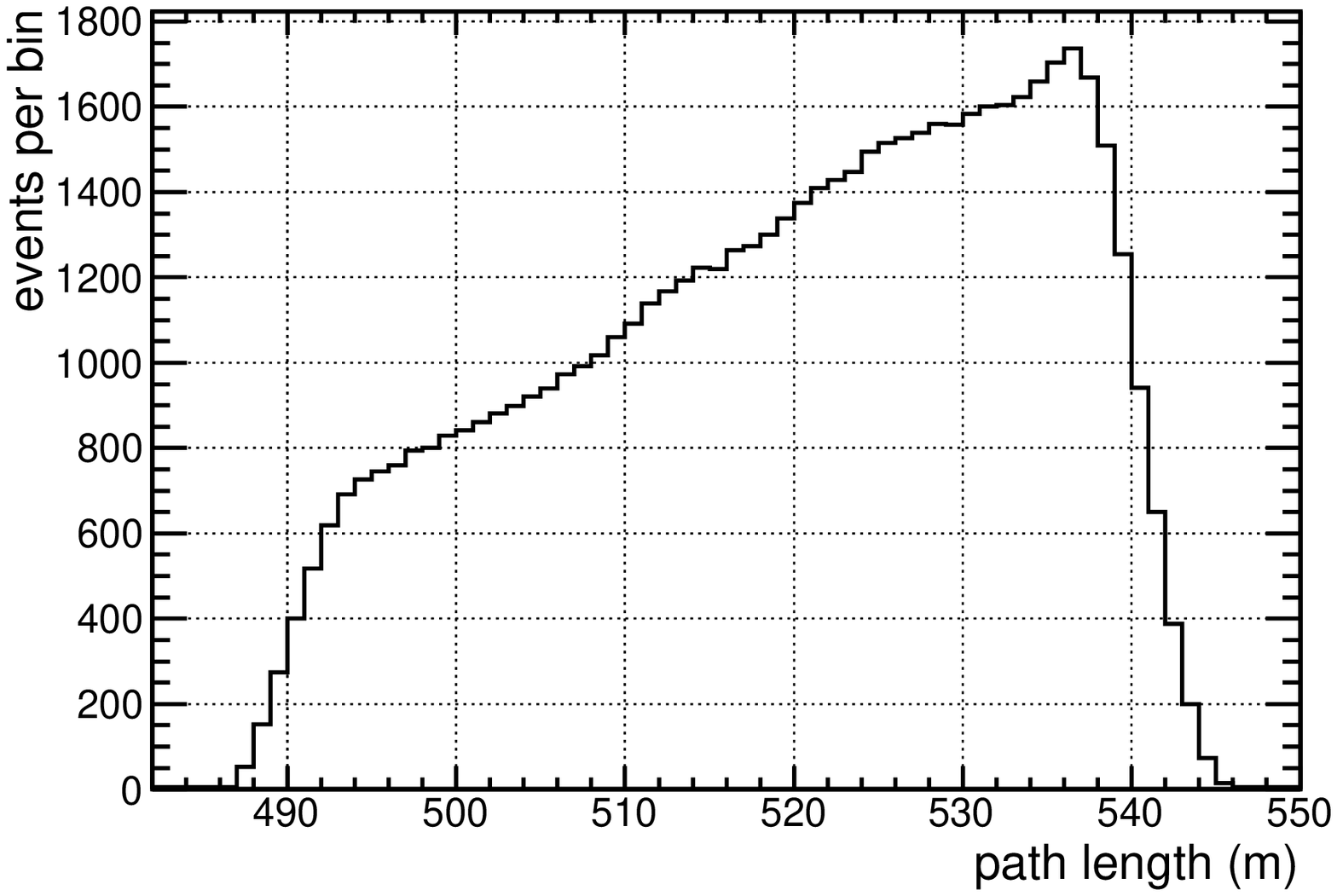}}
\subfigure[~SciBooNE (anti)neutrino path lengths]{\includegraphics[width=0.95\columnwidth]{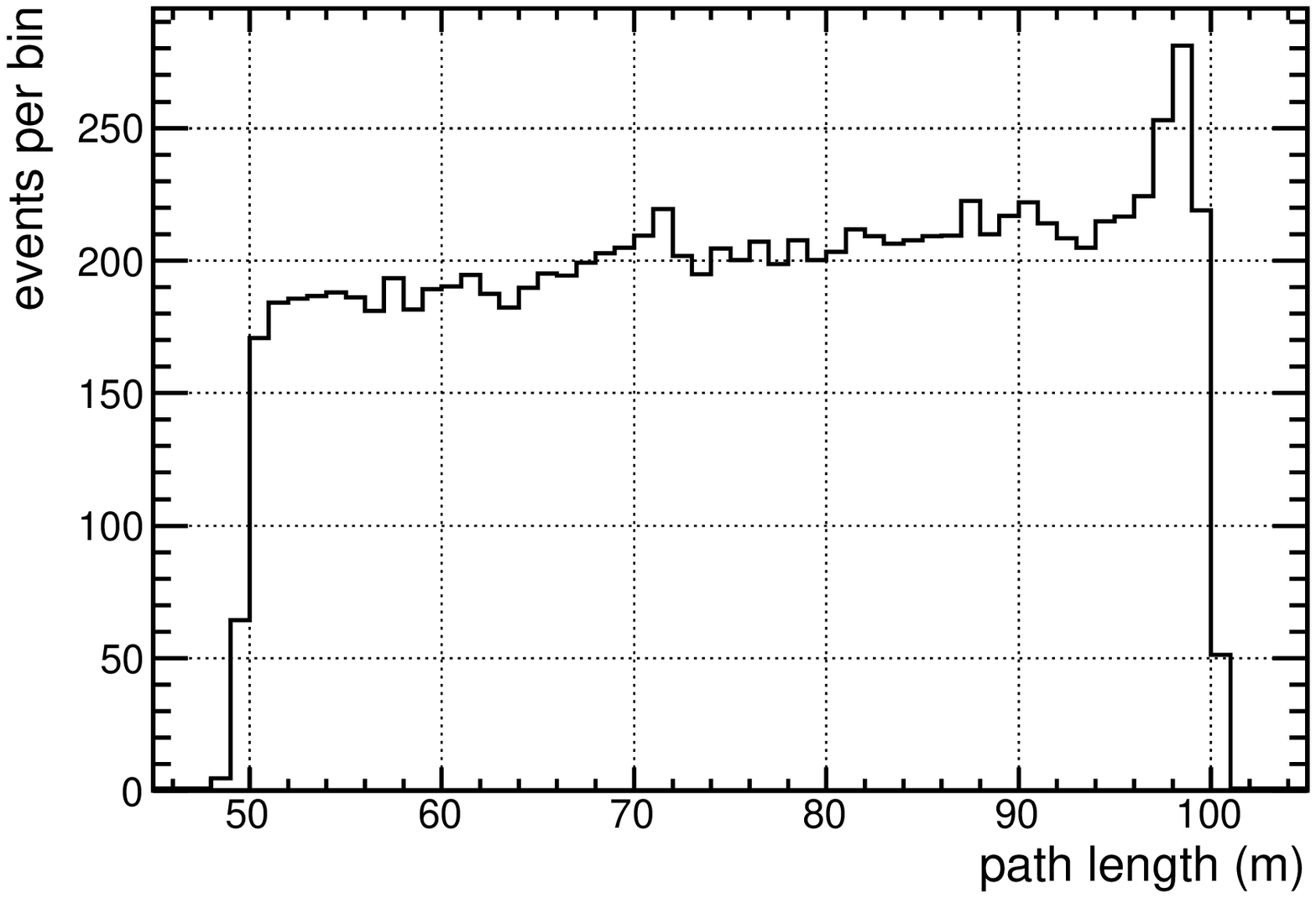}}
\caption{Antineutrino and neutrino path lengths for MiniBooNE and SciBooNE events from point of production to interaction in detector, as predicted by the MC simulation.}
\label{fig:mbsb_pathlength}
\end{center}
\end{figure}

Figure~\ref{fig:mbsb_pathlength} shows the distribution of the combined antineutrino and neutrino propagation distances, from production in the decay tunnel to interaction in SciBooNE or MiniBooNE.

\section{Systematic Uncertainties}\label{sec:sysuncertain}

Beam and cross section uncertainties are calculated for both MiniBooNE and SciBooNE using the multisim method~\cite{Mahn:2009zz}.  In this procedure, groups of correlated simulation parameters associated with beam production and cross section modeling uncertainties are sampled according to their covariance matrices. The parameters for each source of uncertainty ($\pi^{\pm}$, $K^{+}$, etc.) are sampled 1000 times to obtain sufficient statistics. Each MC event in MiniBooNE and SciBooNE is reweighted based on these varied parameters forming 1000 new MC predictions of the $E_{\nu}^{QE}$ distribution in both detectors.  Covariance matrices, in bins of $E_{\nu}^{QE}$, are then computed for each source of uncertainty by comparing these 1000 new MC predictions to the default MC prediction. The procedure takes care of the correlation of beam production and cross section uncertainties between MiniBooNE and SciBooNE.  Cross section and nuclear model uncertainties for $\nu$ and $\bar{\nu}$ events are treated as uncorrelated due to the poor understanding of differences between $\nu$ and $\bar{\nu}$ interactions in nuclear modeling. Some detector specific uncertainties are calculated using the unisim method~\cite{Mahn:2009zz}, where uncorrelated detector specific uncertainties are varied up or down by 1 standard deviation.

\subsection{Beam Uncertainties}

Uncertainties in the delivery of the primary proton beam to the beryllium target, the primary beam optics, secondary hadron production in proton-beryllium interactions, hadronic interactions in the target and horn, and the horn magnetic field, are included in the beam multisims.  Uncertainties in the magnetic field horn current, skin effect of the horn, and secondary nucleon and pion interactions in the Be target and Al horn are obtained from previous MiniBooNE analyses~\cite{AguilarArevalo:2008yp}.

The normalization of the neutrino component in the antineutrino beam was adjusted based on direct measurements in MiniBooNE~\cite{PhysRevD.84.072005, Grange:2011zi}.  The beam fraction of $\nu_\mu$ in the antineutrino beam was determined using three methods: a pure data sample of $\nu_{\mu}$ events from CC$1\pi$ interactions, differences in Michel electron rates between final state $\mu^-$ and $\mu^+$ from $\nu_\mu$ and $\bar\nu_{\mu}$ interactions, respectively, due to $\mu^-$ capture on carbon, and angular distribution differences between final state $\mu^-$ and $\mu^+$ from $\nu_\mu$ and $\bar\nu_{\mu}$ interactions, respectively. Averaging these three methods, the $\pi^+$ production in the beam MC was scaled by a factor of 0.78 and given a $12.8\%$ normalization uncertainty. Uncertainties on the production of $\pi^-$ from the initial \emph{p}-Be interaction are calculated using spline fits to data from the HARP experiment~\cite{AguilarArevalo:2008yp}. An updated $K^+$ production simulation with reduced uncertainties for the initial \emph{p}-Be interaction is used. This update is based on a new Feynman scaling fit~\cite{PhysRevD.84.114021} to recent SciBooNE measurements~\cite{PhysRevD.84.012009}. The $K^0$ production uncertainties for the initial p-Be interaction are from the Sanford-Wang parametrization covariance matrix~\cite{AguilarArevalo:2008yp}. $K^-$ production is estimated using the MARS hadronic interaction package~\cite{Mokhov:1998kc} due to the scarcity of production measurements in the relevant kinematic regions. $K^-$ production cross section uncertainties from the initial \emph{p}-Be interaction are given a conservative $100\%$ normalization uncertainty.

\subsection{Cross Section Uncertainties}

CCQE cross sections on carbon are calculated assuming an RFG model with parameters $M_{A}$ (axial mass) $= 1.35$ and $\kappa$ (Pauli blocking factor) $= 1.007$.  An additional correction, as a function of $Q^2$, is applied to background CC$1\pi$ interaction events in MC~\cite{PhysRevD.81.092005}. The uncertainties in $M_A$ and $\kappa$ for CCQE events on carbon are based on the statistical uncertainties of the MiniBooNE neutrino mode measurement~\cite{PhysRevD.81.092005}, to avoid double counting systematic uncertainties accounted for in this analysis as detailed in this section.

Since the purpose of the $Q^2$ correction in the MiniBooNE neutrino mode measurement~\cite{PhysRevD.81.092005} is to match the background CC$1\pi$ interaction events in MC to a selected data sample comprising mainly of CC$1\pi$ interaction events, there is no uncertainty placed on $M_{A}$ for $\nu_\mu$ CC$1\pi$ interaction events.  However, for $\bar\nu_{\mu}$ CC$1\pi$ interaction events, the $M_A$-resonant and coherent 1$\pi$ uncertainties are not constrained by the MiniBooNE neutrino mode measurement and are not reduced. The values and uncertainties of $M_{A}$ for CC coherent $\pi$ interactions, $M_{A}$ for multi $\pi$ interactions, Fermi surface momentum ($p_F$), and NC axial vector isoscalar contribution ($\Delta$s) are identical to previous MiniBooNE and SciBooNE measurements~\cite{Nakajima:2010fp,Mahn:2011ea}. The uncertainties for pion absorption, pion inelastic scattering, and pionless $\Delta$ decay in the target nucleus ($\pm 25\%$, $\pm 30\%$, and $\pm 100\%$, respectively) are treated in the same way as in a previous measurement~\cite{Mahn:2011ea}, however they are treated as uncorrelated between MiniBooNE and SciBooNE (unlike all other cross section uncertainties). Both the $\nu$ and $\bar{\nu}$ $M_A$ values and their uncertainties for quasielastic interactions on hydrogen are based on the latest deuterium measurements~\cite{Bodek:2007vi}.

Additional systematic uncertainties are added to account for limitations of the RFG model. Such limitations include the absence of processes such as meson exchange currents and multinucleon knockout events~\cite{Amaro:2011aa,Bodek:2011ps,Nieves:2011yp,Martini:2009uj}. A 10\% normalization uncertainty is assigned to both $\nu$ and $\bar{\nu}$ CCQE interactions on carbon to cover the difference between data and prediction in the MiniBooNE $\nu_\mu$ CCQE measurement.  An additional 40\% normalization uncertainty is placed on $\bar{\nu}$ CCQE interactions on carbon to cover the discrepancy between the RFG model prediction for $\bar{\nu}$ and recent nuclear models~\cite{Amaro:2011aa,Bodek:2011ps,Nieves:2011yp,Martini:2009uj}. An additional 10\% normalization uncertainty is added to non-CCQE $\bar{\nu}$ interactions on carbon to account for the limitations of the RFG model for those type of events.

The full list of beam and cross section parameters for MC simulation and its associated systematic uncertainties are shown in Table~\ref{tab:sys_uncert}.

\subsection{Detector Uncertainties}

Uncertainties associated with the MiniBooNE detector include light propagation, attenuation, and scattering in the detector as well as PMT response. The optical model for light propagation in the detector~\cite{BrownIEEE:2004} uses 35 parameters for properties such as refractive index, attenuation length, scintillation strength, etc.  These parameters are tuned to non-MiniBooNE measurements as well as MiniBooNE internal data.  Over 100 separate MC data sets were created based on variations in these parameters.  In a manner similar to the multisim method, these results were used to compute the optical model error matrix in bins of reconstructed antineutrino energy.  To estimate the impact of uncertainties in PMT response, independent MC data sets based on variations in the discriminator threshold, or the PMT charge-time correlations, were created and compared to default MC.  Based on comparisons with external data~\cite{Ashery:1981tq,Jones:1993ps,Ransome:1992bx} and the output of GCALOR, an uncertainty of $35\%$ is assigned to pion absorption and $50\%$ is assigned to charge exchange in the detector medium.  This is distinct from the uncertainty on pion absorption and charge exchange inside the nucleus.

Uncertainties associated with the SciBooNE detector include uncertainties in the muon energy loss in the scintillator and iron, light attenuation in the wavelength shifting fibers, and PMT response; see Ref.~\cite{Nakajima:2010fp}. The crosstalk of the MA-PMT was measured to be 3.15\% for adjacent channels with an absolute error of 0.4\%~\cite{Hiraide:2008eu}. The single photoelectron resolution of the MA-PMT is set to 50\% in the simulation, and the absolute error is estimated to be $\pm 20$\%. Birk's constant for quenching in the SciBar scintillator was measured to be $0.0208 \pm 0.0023$ cm/MeV~\cite{Hiraide:2008eu}. The conversion factors for ADC counts to photoelectrons were measured for all 14336 MA-PMT channels in SciBar. The measurement uncertainty was at the 20\% level. The threshold for hits to be used in SciBar track reconstruction is 2.5 photoelectrons; this threshold is varied by $\pm$20\% to evaluate the systematic error for SciBar track reconstruction. The TDC dead time is set to 55 ns in the MC simulation, with the error estimated to be $\pm$20~ns~\cite{Hiraide:2009zz}.

The reconstruction uncertainties consist of antineutrino energy reconstruction uncertainties and muon track misidentification uncertainties. For antineutrino energy reconstruction uncertainties, the densities of SciBar, EC, and MRD are varied independently within their measured uncertainties of $\pm 3\%$, $\pm 10\%$, and $\pm 3\%$, respectively. Misidentified muons stem mainly from proton tracks created through NC interactions, which are given a conservative $\pm 20\%$ normalization uncertainty. A conservative $\pm 20\%$ normalization uncertainty is applied for the MC simulated background of neutrino and antineutrino events initially interacting outside the SciBooNE detector that pass the selection criteria.  A conservative $\pm 20\%$ normalization uncertainty is applied for the MC simulated background of neutrino and antineutrino events initially interacting in the EC/MRD detector that pass the selection criteria.

\begin{table*}
\caption{\label{tab:sys_uncert} Summary of beam and cross section parameters for MC simulation with its associated systematic uncertainties.}
\begin{ruledtabular}
\begin{tabular}{lc}
Beam & Uncertainty \\
\hline
$\pi^+$ production in antineutrino beam (from WS neutrino background) & 12.8\% normalization uncertainty~\cite{PhysRevD.84.072005} \\
$\pi^-$ production from p-Be interaction & Spline fit to HARP data \\
$K^+$ production from p-Be interaction & Table IX in Ref.~\cite{PhysRevD.84.114021} \\
$K^0$ production from p-Be interaction & Table IX in Ref.~\cite{AguilarArevalo:2008yp} \\
$K^-$ production from p-Be interaction & 100\% normalization uncertainty \\
Nucleon and pion interaction in Be/Al & Table XIII in Ref.~\cite{AguilarArevalo:2008yp} \\
Horn current & $\pm$1 kA \\
Horn skin effect & Horn skin depth, $\pm$1.4 mm \\
\hline
Cross Sections & Uncertainty \\
\hline
CCQE $M_{A}$ on carbon target & 1.35 $\pm$0.07~GeV \\
$\kappa$ & 1.007 $\pm$0.005 \\
CCQE $M_{A}$ on hydrogen target & 1.014 $\pm$0.014~GeV \\
CC resonant $\pi$ $M_{A}$ & 1.1 $\pm$0.275~GeV \footnotemark[1]\\
CC coherent $\pi$ $M_{A}$ & 1.03 $\pm$0.275~GeV \footnotemark[1]\\
CC multi $\pi$ $M_{A}$ & 1.3 $\pm$0.52~GeV \\
$E_{B}$ & $\pm$9~MeV \\
$p_F$ & 220 $\pm$30~MeV/c \\
$\Delta$s & 0.0 $\pm$0.1 \\
CCQE on carbon & $\pm 10\%$ norm error \\
CCQE on carbon $\left( \bar\nu_{\mu} \right)$ only & $\pm 40\%$ norm error \\
non-CCQE on carbon $\left( \bar\nu_{\mu} \right)$ only & $\pm 10\%$ norm error \\
$\pi$ absorption in nucleus & 25\% \\
$\pi$ inelastic scattering & 30\% \\
$\pi$-less $\Delta$ decay & 100\% \\
\end{tabular}
\footnotetext[1]{This uncertainty is not applied to $\nu_\mu$ CC$1\pi$ events that are $Q^{2}$ corrected.}
\end{ruledtabular}
\end{table*}

\subsection{Error Matrix}\label{subsec:error_matrix}

All of the MiniBooNE uncertainties, the SciBooNE uncertainties, and the correlations between them are expressed in the total error matrix, $M$, a 42 $\times$ 42 covariance matrix in MiniBooNE and SciBooNE reconstructed antineutrino energy bins defined as
\begin{eqnarray}
M = \left( {\begin{array}{*{20}c}
   {M^{{\rm{MB\mbox{-}SB}}} } & {M^{{\rm{SB\mbox{-}SB}}} }  \\
   {M^{{\rm{MB\mbox{-}MB}}} } & {M^{{\rm{SB\mbox{-}MB}}} }  \\
\end{array}} \right)
\label{eq:error_matrix}
\end{eqnarray}
where
\begin{eqnarray}
M_{i,j}^{{\rm{X}}}  &=& \hat M_{i,j;(RS,RS)}^{{\rm{X}}} N_i^{{\rm{Y}} \; RS} N_j^{{\rm{Z}} \; RS} \nonumber \\
                        && {}+ \hat M_{i,j;(WS,WS)}^{{\rm{X}}} N_i^{{\rm{Y}} \; WS} N_j^{{\rm{Z}} \; WS} \nonumber \\
                        && {}+ \hat M_{i,j;(RS,WS)}^{{\rm{X}}} N_i^{{\rm{Y}} \; RS} N_j^{{\rm{Z}} \; WS} \nonumber \\
                        && {}+ \hat M_{i,j;(WS,RS)}^{{\rm{X}}} N_i^{{\rm{Y}} \; WS} N_j^{{\rm{Z}} \; RS} \nonumber \\
                        && {}+ M_ {i,j}^{{\rm{X}} \; stat}
\end{eqnarray}
are the bin to bin covariance elements of the full error matrix. X denotes the type of correlation with Y and Z denoting the type of bins (either MiniBooNE or SciBooNE) associated with X. For MiniBooNE to MiniBooNE correlations, X=MB-MB, Y=MB, Z=MB. For SciBooNE to SciBooNE correlations, X=SB-SB, Y=SB, Z=SB. For MiniBooNE to SciBooNE correlations, X=MB-SB, Y=MB, Z=SB. For SciBooNE to MiniBooNE correlations, X=SB-MB, Y=SB, Z=MB. $N_i^{{\rm{Y}} \; RS}$ ($N_j^{{\rm{Z}} \; RS}$) and $N_i^{{\rm{Y}} \; WS}$ ($N_j^{{\rm{Z}} \; WS}$) are the number of RS and WS events for bin type Y (bin type Z) in reconstructed antineutrino energy bin $i$ (bin $j$), respectively. $\hat M_{i,j;(RS,RS)}^{{\rm{X}}}$ are the elements of the RS to RS correlated fractional error matrix for correlation type X defined as:
\begin{equation}
\hat M_{i,j;(RS,RS)}^{{\rm{X}}}  = \frac{{M_{i,j;(RS,RS)}^{{\rm{X}}}}}{{N_i^{{\rm{Y}} \; RS} N_j^{{\rm{Z}} \; RS} }}
\end{equation}
where $M_{i,j;(RS,RS)}^{{\rm{X}}}$ is the full RS to RS reconstructed antineutrino energy bin covariance for correlation type X. $\hat M_{i,j;(WS,WS)}^{{\rm{X}}}$, $\hat M_{i,j;(RS,WS)}^{{\rm{X}}}$, and $\hat M_{i,j;(WS,RS)}^{{\rm{X}}}$ are similarly defined fractional error matrices for correlation type X with different RS and WS correlations. $M^{{\rm{X}} \; stat}$ is the statistical covariance matrix in reconstructed antineutrino energy bins for correlation type X (only SB-SB and MB-MB have nonzero elements).

The decomposition and reconstruction of the full error matrix $M$ to and from the fractional error matrices allows the error matrix to be updated based on different MC predictions, as a function of the oscillation parameters in the physics parameter space.

\begin{figure}[htbn!]
\begin{center}
\includegraphics[width=0.95\columnwidth]{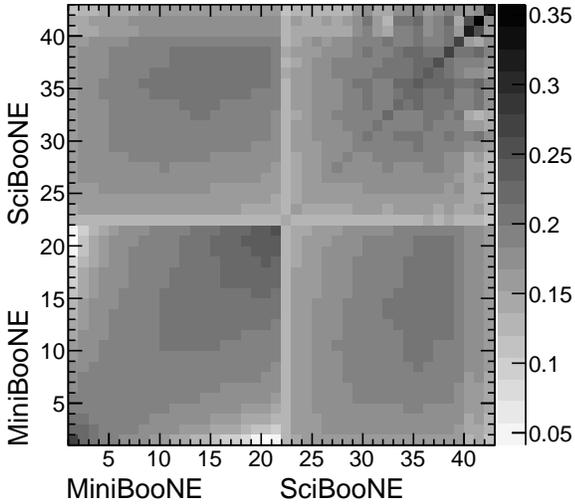}
\caption{Bin-wise square root of the total (statistical and systematic errors combined) fractional error matrix $\sqrt {\hat M_{ij}} = \sqrt {M _{ij}}  / \sqrt {N_i N_j}$, where $M _{ij}$ is the total error matrix and $N_i$ ($N_j$) is the MC prediction for reconstructed antineutrino energy bin $i$ ($j$).  Bins 1 through 21 are MiniBooNE, bins 22 through 42 are SciBooNE.}
\label{fig:frac_err}
\end{center}
\end{figure}

\begin{figure}[htbn!]
\begin{center}
\includegraphics[width=0.95\columnwidth]{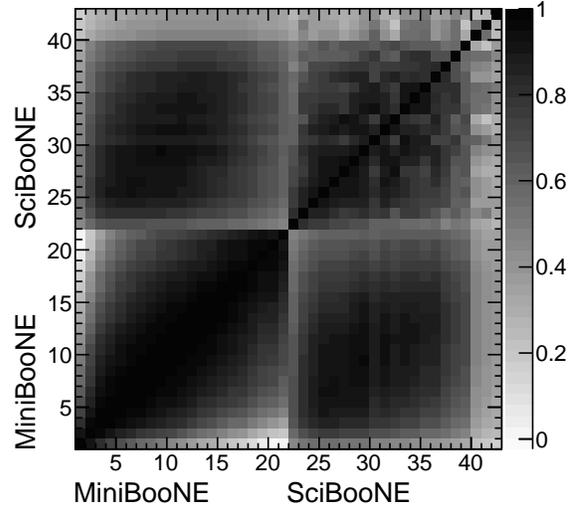}
\caption{Correlation coefficients of the total (statistical and systematic errors combined) error matrix $\left({\rho_{ij}} = M _{ij}  / (\sigma _{ii}\sigma _{jj})  \right)$.  Bins 1 through 21 are MiniBooNE, bins 22 through 42 are SciBooNE.  No bins are anticorrelated.}
\label{fig:total_corr}
\end{center}
\end{figure}

For illustrative purposes, Fig.~\ref{fig:frac_err} shows the square roots of the elements of the total fractional error matrix, $\sqrt {\hat M_{ij}} = \sqrt {M _{ij}}  / \sqrt {N_i N_j}$, where $M _{ij}$ are the elements of the total error matrix and $N_i$ ($N_j$) is the MC prediction for reconstructed antineutrino energy bin $i$ ($j$). Figure~\ref{fig:total_corr} shows the correlation coefficients of the total error matrix in reconstructed antineutrino energy bins.

\begin{figure*}[htbn!]
\begin{center}
\subfigure[~MiniBooNE RS and WS events with systematic]{\includegraphics[width=1.0\columnwidth]{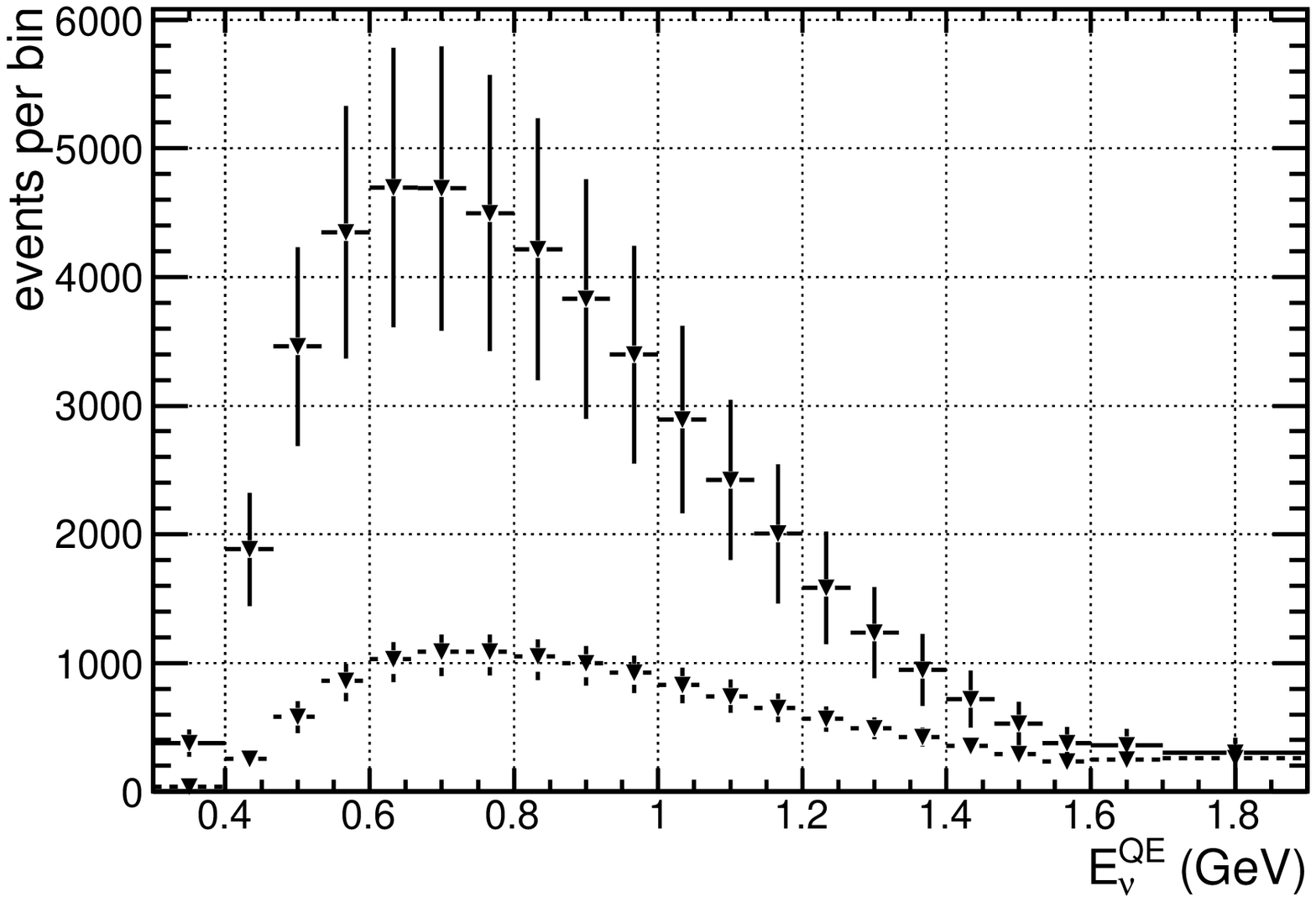}}
\subfigure[~SciBooNE RS and WS events with systematics]{\includegraphics[width=1.0\columnwidth]{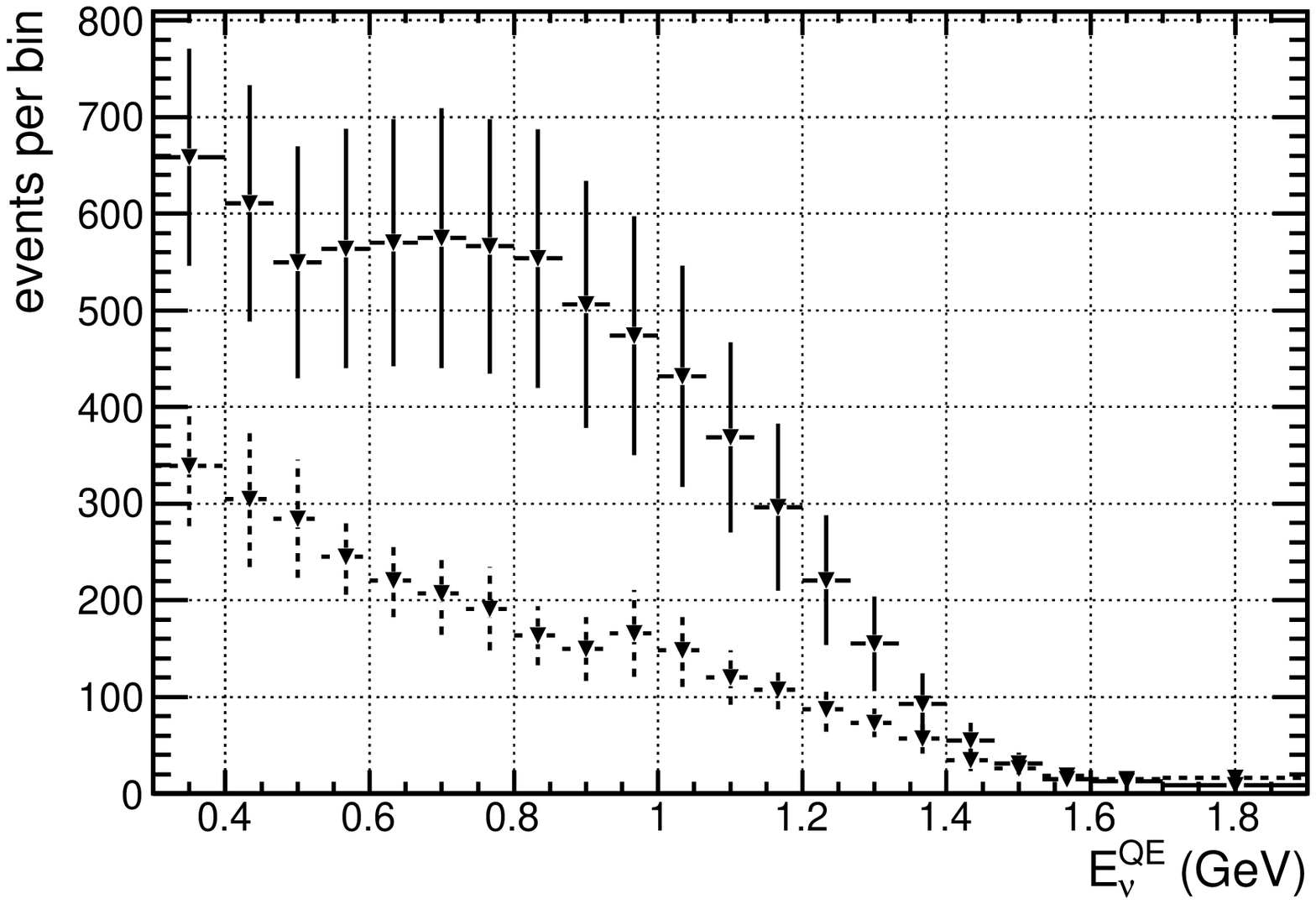}}
\caption{RS (solid line) and WS (dashed line) contributions to the default MC prediction for MiniBooNE and SciBooNE reconstructed antineutrino and neutrino energy ($E_\nu^{QE}$) distributions.  Error bars are the systematic uncertainties from the diagonals of the error matrix $\left( \sigma_{ii}=\sqrt{M_{ii}} \right)$ and do not account for correlations.}
\label{fig:MB_enu_CV}
\end{center}
\end{figure*}

Figure~\ref{fig:MB_enu_CV} shows the MiniBooNE and SciBooNE default MC $E_\nu^{QE}$ predictions for RS and WS events with error bars corresponding to the $\sqrt{M_{ii}}$ values of the error matrix diagonal elements.

\section{Analysis Methodology}\label{sec:method}

Oscillation predictions are based on a two-antineutrino oscillation model, where the oscillation survival probability for a $\bar{\nu}_{\mu}$ in the beam is given by
\begin{equation}
P\left(\bar{\nu}_{\mu} \rightarrow \bar{\nu}_x \right)  = 1 - \sin ^2 2\theta \sin ^2 \frac{{1.27\Delta m^2 L}}{E}.
\end{equation}
$L$ is the path length in kilometers, $E$ is the antineutrino energy in GeV, $\theta$ is the mixing angle, and $\Delta m^2$ is the difference in the squares of the masses of two different mass eigenstates.

The $\chi^{2}$ statistic is formed,
\begin{equation}
\chi ^2  = \sum\limits_{i,j = 1}^{42} {\left( {D_i  - N_i } \right)\left( {M^{ - 1} } \right)_{ij} \left( {D_j  - N_j } \right)},
\end{equation}
where $(M^{-1})_{ij}$ is the $ij$-th element of the inverse of the error matrix $M$, the covariance matrix in MiniBooNE and SciBooNE $E_\nu^{QE}$ bins described in Eq.~\ref{eq:error_matrix}. $D_{i}$ ($D_{j}$) is the data count in bin $i$ ($j$) and $N_{i}$ ($N_{j}$) is the MC prediction for bin $i$ ($j$), in MiniBooNE and SciBooNE $E_\nu^{QE}$ bins. $N_{i}$ is the sum of neutrino (WS) and antineutrino (RS) events in the $i$th bin:
\begin{equation}
N_i  = N_i^{RS} \left( {\Delta m^2 ,\sin ^2 2\theta } \right) + N_i^{WS}.
\label{eq:x_i}
\end{equation}
As shown in Eq.~\ref{eq:x_i}, only the predicted RS event rate depends on the oscillation parameters, $\Delta m^2$ and $\sin ^2 2\theta$, for this two-antineutrino oscillation model.  The WS flux is assumed to not oscillate. The index runs from 1 to 42 (21 MiniBooNE $E_\nu^{QE}$ bins and 21 SciBooNE $E_\nu^{QE}$ bins).  For the physics analysis fitting, a $\Delta\chi^2$ test statistic is used as defined by
\begin{eqnarray}
\Delta \chi ^2  &=& \chi ^2 \left( {N(\theta _{{\rm{phys}}} ),M(\theta _{{\rm{phys}}} )} \right)  \nonumber \\
                         && {}- \chi ^2 \left( {N(\theta _{{\rm{BF}}} ),M(\theta _{{\rm{BF}}} )} \right)
\end{eqnarray}
where $\theta _{{\rm{BF}}}$ refers to the oscillation parameters at the best fit point and $\theta _{{\rm{phys}}}$ refers to the oscillation parameters at a given test point.

The method of Feldman and Cousins~\cite{PhysRevD.57.3873} is used to determine the $\Delta \chi^2$ value at each point that corresponds to a certain confidence level of acceptance or rejection. To obtain the 90\% confidence level exclusion region for $\bar{\nu}_{\mu}$ disappearance, a $\Delta\chi^2$ distribution is formed for each point $\theta _{{\rm{phys}}}$ in parameter space using many iterations of generated fake data at that $\theta _{{\rm{phys}}}$. The $\Delta\chi^2$ value from actual data at each $\theta _{{\rm{phys}}}$ is then compared to the fake data $\Delta\chi^2$ distribution at each $\theta _{{\rm{phys}}}$. If the $\Delta\chi^2$ value from actual data is larger than 90\% of the all the fake data $\Delta\chi^2$ values at a $\theta _{{\rm{phys}}}$ point, then the $\theta _{{\rm{phys}}}$ point in parameter space is excluded at 90\% confidence level. The aggregation of all the excluded 90\% confidence level $\theta _{{\rm{phys}}}$ points forms the 90\% confidence level exclusion region.

The full error matrix is used to create the fake data for the Feldman and Cousins tests.  First, a Cholesky decomposition is performed on the error matrix $M$:
\begin{eqnarray}
M = LL^* ,
\end{eqnarray}
where $L$ is a lower triangular matrix and $L^*$ is the conjugate transpose of $L$.  Then, a vector $u$ is created, where each of the $n$ elements, 42 in total, of $u$ are drawn from a Gaussian distribution with mean 0 and variance 1.  A fluctuated fake data histogram is given by
\begin{eqnarray}
N_{{\rm{fake}}}  = N_{default} \left( {\theta _{{\rm{phys}}} } \right) + Lu,
\end{eqnarray}
where $N_{default}$ is the default Monte Carlo prediction assuming an oscillation signal with oscillation parameters at point $\theta _{{\rm{phys}}}$.

\section{Results}\label{sec:results}

\begin{figure*}[htbn!]
\begin{center}
\subfigure[~MiniBooNE]{\includegraphics[width=1.0\columnwidth]{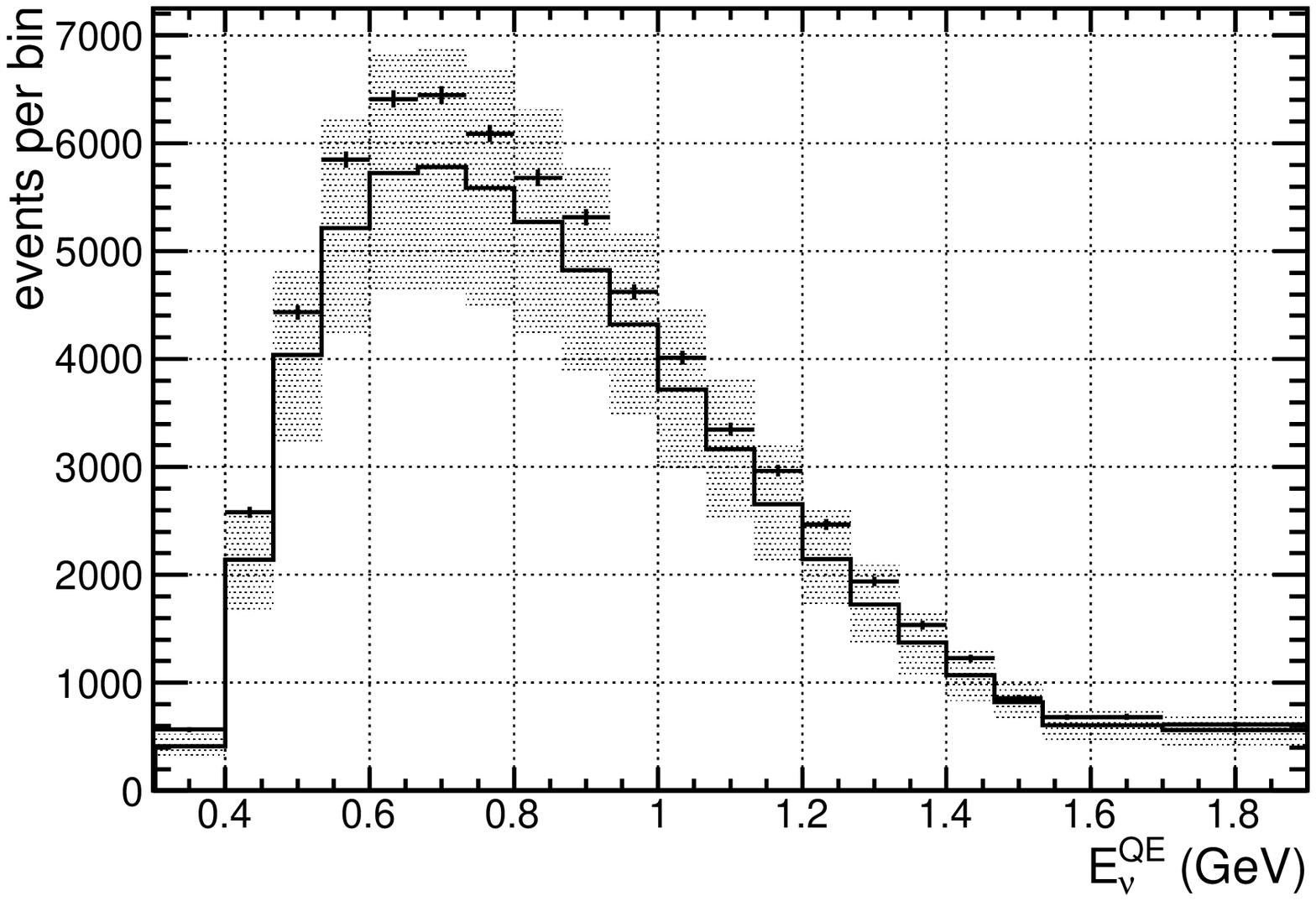}}
\subfigure[~SciBooNE]{\includegraphics[width=1.0\columnwidth]{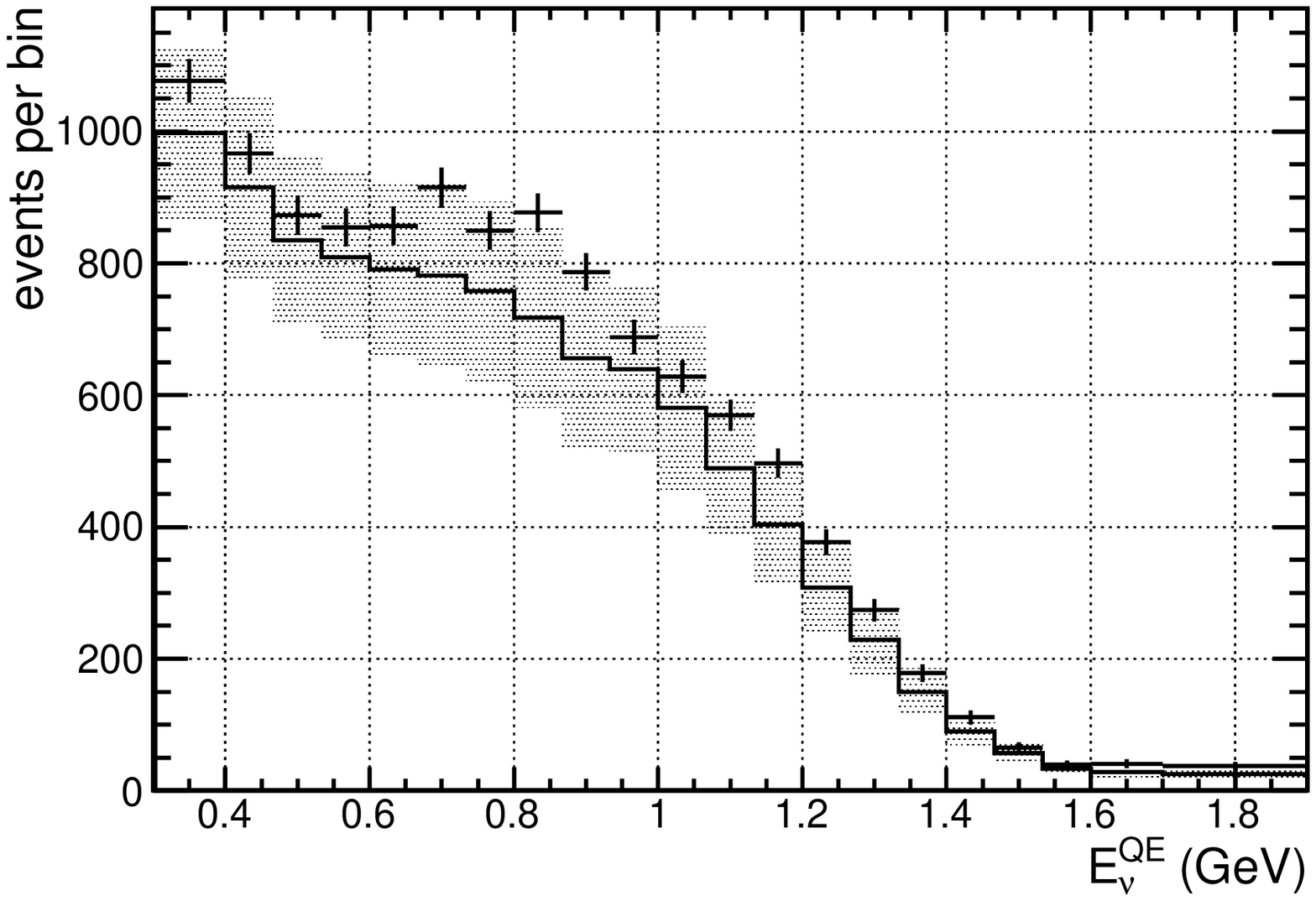}}
\caption{Reconstructed antineutrino energy ($E_\nu^{QE}$) distribution for data events, compared to Monte Carlo predictions, for MiniBooNE and SciBooNE.  Vertical error bars on data are statistical uncertainty.  Shaded error band around simulation is the systematic uncertainty computed as the square roots of the diagonals of the total error matrix.}
\label{fig:data}
\end{center}
\end{figure*}

Figure~\ref{fig:data} shows the observed event distributions, in reconstructed antineutrino energy, for MiniBooNE and SciBooNE.  The systematic uncertainty shown for the MC predictions was computed as just the square roots of the diagonals of the total error matrix without correlations.

Table~\ref{tab:summary} lists the event counts in each bin, for data and for MC predictions.  The listed uncertainties are based on the square roots of the diagonals of the total error matrix without correlations. (The reported SciBooNE data has fractional counts due to the manner in which the cosmic ray background is subtracted.)

\begin{table*}[htbn!]
\caption{Observed event counts for each MiniBooNE and SciBooNE data bin, MC predictions, and uncertainty.}
\begin{tabular}{cccccccc}
\hline\hline
Bin range (MeV) &MB data &MB MC  &MB error $\pm$        &SB data\footnotemark[2]    &SB cosmic    &SB MC &SB error $\pm$         \\
\hline
300-400&565&413.5&111.0&1077.0&21.0&997.3&136.8\\
400-467&2577&2139.2&464.8&966.8&89.2&915.6&141.9\\
467-533&4433&4039.9&802.2&872.8&85.2&834.4&132.1\\
533-600&5849&5211.0&1005.7&854.4&72.6&809.4&132.2\\
600-667&6411&5725.6&1108.7&856.8&59.2&790.6&137.3\\
667-733&6445&5778.3&1130.3&915.0&51.0&781.9&144.3\\
733-800&6090&5586.8&1096.9&849.8&52.2&757.3&139.5\\
800-867&5678&5268.3&1044.8&876.6&43.4&717.1&138.8\\
867-933&5314&4826.2&951.8&787.0&39.0&655.8&138.0\\
933-1000&4624&4319.6&865.1&688.0&35.0&639.7&129.6\\
1000-1067&4015&3720.3&747.2&628.0&29.0&580.2&125.4\\
1067-1133&3349&3163.6&642.1&569.6&28.4&488.7&105.8\\
1133-1200&2965&2655.9&554.3&496.6&21.4&403.9&92.2\\
1200-1267&2464&2147.2&453.0&377.0&23.0&308.4&74.6\\
1267-1333&1937&1726.4&367.8&273.6&22.4&228.4&53.6\\
1333-1400&1534&1372.0&297.9&178.6&18.4&150.0&37.2\\
1400-1467&1227&1073.3&238.1&111.2&18.8&89.4&23.9\\
1467-1533&859&820.5&187.7&65.4&17.6&57.1&15.0\\
1533-1600&679&607.2&145.8&39.0&17.0&33.1&10.4\\
1600-1700&684&607.2&149.1&40.8&28.2&27.6&9.9\\
1700-1900&610&560.1&144.5&37.6&39.4&24.8&7.8\\
\hline\hline
\end{tabular}
\footnotetext[2]{The SB data has its SB cosmic data background removed.}
\label{tab:summary}
\end{table*}

\begin{figure}[htbn!]
\begin{center}
\includegraphics[width=1.0\columnwidth]{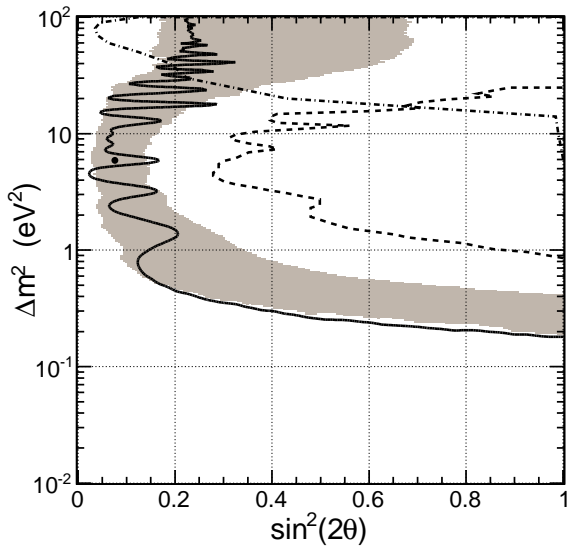}
\caption{$90\%$ C.L. exclusion region (solid line) and best fit point for the MiniBooNE-only $\bar\nu_{\mu}$ disappearance analysis.  Also shown is the $90\%$ C.L. result from the 2009 MiniBooNE disappearance analysis~\cite{PhysRevLett.103.061802} (dashed line) and the Chicago-Columbia-Fermilab-Rochester (CCFR) experiment~\cite{Stockdale:1984cg} (dot-dashed line).  The expected $90\%$ C.L. sensitivity band from fake data studies is also shown (shaded region); $1\sigma$ ($68\%$) of fake data tests, where the fake data had statistical and systematic fluctuations but no oscillation signal, had $90\%$ C.L. limit curves in this shaded region.}
\label{fig:mb_results}
\end{center}
\end{figure}

A MiniBooNE-only disappearance analysis is included to give a sense of what the sensitivity would be without the inclusion of SciBooNE data. Figure~\ref{fig:mb_results} shows the $90\%$ C.L. exclusion region and best fit point for the MiniBooNE-only $\bar\nu_{\mu}$ disappearance analysis, completed using the same methodology as the joint disappearance analysis except with the exclusion of SciBooNE data, SciBooNE MC prediction, and SciBooNE error matrix uncertainties in the $\chi^{2}$ statistic.  The best fit point is $\Delta m^2 = 5.9$ ${\mathrm{eV}^{2}}$, $\sin^2 2\theta = 0.076$.   At the best fit point, $\chi^{2} = 25.7$ (probability $12.4\%$).  For the null oscillation hypothesis, $\chi^{2} = 28.3$ (probability $13.7\%$).  With $\Delta\chi^{2} = 2.6$, the null oscillation hypothesis is excluded at $52.4\%$ C.L.

\begin{figure}[htbn!]
\begin{center}
\subfigure[~Linear scale on x-axis.]{\includegraphics[width=1.0\columnwidth]{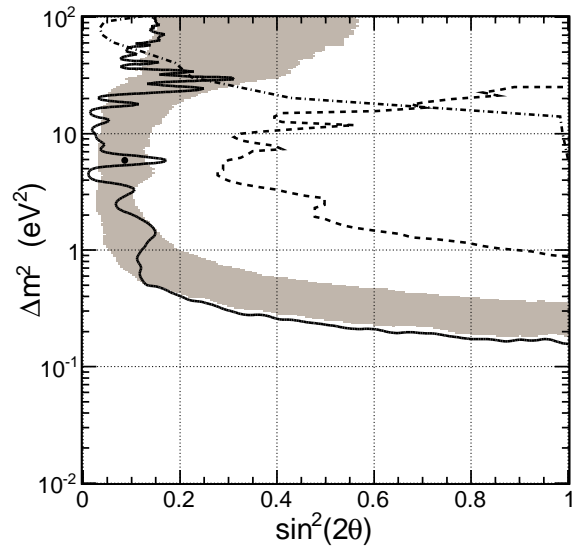}}
\subfigure[~Log scale on x-axis.]{\includegraphics[width=1.0\columnwidth]{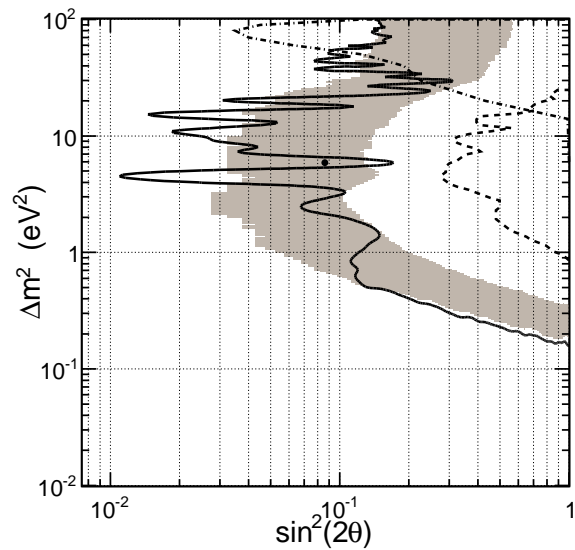}}
\caption{$90\%$ C.L. exclusion region (solid line) and best fit point for the joint MiniBooNE, SciBooNE $\bar\nu_{\mu}$ disappearance analysis.  Also shown is the $90\%$ C.L. result from the 2009 MiniBooNE disappearance analysis~\cite{PhysRevLett.103.061802} (dashed line) and the CCFR experiment~\cite{Stockdale:1984cg} (dot-dashed line).  The expected $90\%$ C.L. sensitivity band from fake data studies is also shown (shaded region); $1\sigma$ ($68\%$) of fake data tests, where the fake data had statistical and systematic fluctuations but no oscillation signal, had $90\%$ C.L. limit curves in this shaded region.}
\label{fig:joint_results}
\end{center}
\end{figure}

Figure~\ref{fig:joint_results} shows the $\bar\nu_{\mu}$ disappearance limit for the joint disappearance analysis. For $\Delta m^2 = 1$ ${\mathrm{eV}^{2}}$ and $\Delta m^2 = 10$ ${\mathrm{eV}^{2}}$, the 90\% C.L. limit for $\sin^2 2\theta$ are at 0.121 and 0.024, respectively. At $\sin^2 2\theta = 1$, the $90\%$ C.L. limit for $\Delta m^2$ is 0.156 ${\mathrm{eV}^{2}}$. The best fit point from the joint analysis is $\Delta m^2 = 5.9$ ${\mathrm{eV}^{2}}$, $\sin^2 2\theta = 0.086$.  At the best fit point, $\chi^{2} = 40.0$ (probability $47.1\%$).  For the null oscillation hypothesis, $\chi^{2} = 43.5$ (probability $41.2\%$).  With $\Delta\chi^{2} = 3.5$, the null oscillation hypothesis is excluded at $81.9\%$ C.L. All probabilities in both the MiniBooNE-only and joint disappearance analyses are based on fake data studies.

\begin{figure*}[htbn!]
\begin{center}
\subfigure[~MiniBooNE]{\includegraphics[width=1.0\columnwidth]{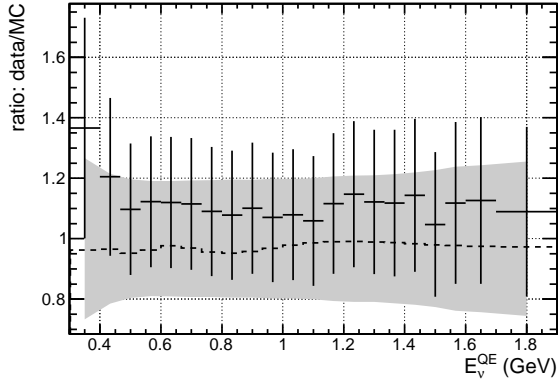}}
\subfigure[~SciBooNE]{\includegraphics[width=1.0\columnwidth]{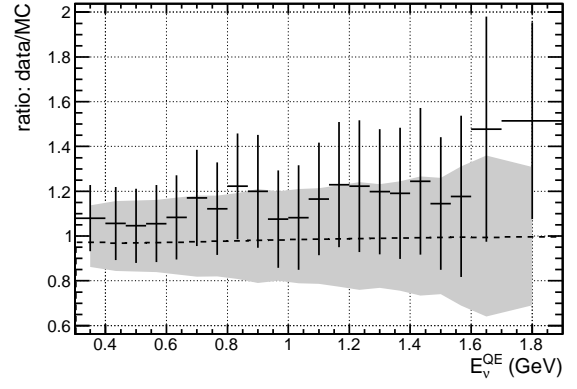}}
\caption{The ratio, with error bars, between data and default MC as a function of reconstructed antineutrino energy ($E_{\nu}^{QE}$). The ratio of best fit signal MC to default MC is also shown (dashed line).  The best fit results from the joint analysis were used to generate the signal MC.  The shaded regions are the $1\sigma$ band from fake data with statistical and systematic fluctuations but no oscillation signal.}
\label{fig:both_ratios}
\end{center}
\end{figure*}

\begin{figure}[htbn!]
\begin{center}
\includegraphics[width=1.0\columnwidth]{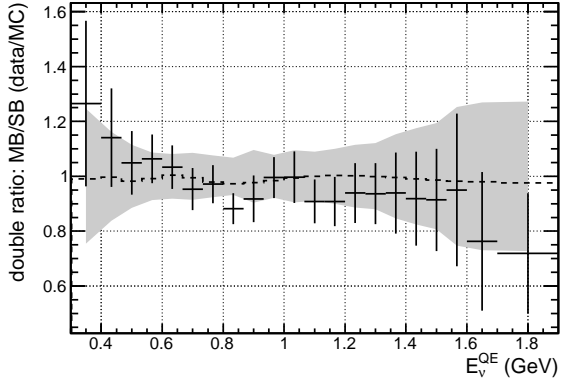}
\caption{The double ratio [Eq.~\ref{dbl_ratio}], with error bars, as a function of reconstructed antineutrino energy ($E_{\nu}^{QE}$).  Some of the MiniBooNE and SciBooNE uncertainties cancel in this double ratio.  The double ratio where the MiniBooNE and SciBooNE signal MC based on the best fit results from the joint analysis are used in placed of data is also shown (dashed line). The shaded region is the $1\sigma$ band from fake data with statistical and systematic fluctuations but no oscillation signal.}
\label{fig:double_ratios}
\end{center}
\end{figure}

Figure~\ref{fig:both_ratios} shows the data to MC ratios for MiniBooNE and SciBooNE, as well as how the best fit signal modifies the MC predictions.  From these ratio plots, it can be seen how the best fit signal improves the shape agreement between data and MC.  Figure~\ref{fig:double_ratios} shows the double ratio
\begin{equation}
\frac{(\frac{\rm{MiniBooNE \; data}}{\rm{MiniBooNE \; default \; MC}})}{(\frac{\rm{SciBooNE \; data}}{\rm{SciBooNE \; default \; MC}})}.
\label{dbl_ratio}
\end{equation}
In Fig.~\ref{fig:double_ratios}, any common normalization difference is removed and the expected result is a value of one. The double ratio result agrees well with the expectation except where statistics are small.

\section{Conclusions}\label{sec:conclusions}

An improved search for $\bar\nu_{\mu}$ disappearance using a two-detector combined MiniBooNE/SciBooNE analysis has been performed. Previous flux and cross section measurements, as well as an increased data set, have enabled a substantial improvement in the sensitivity to $\bar\nu_{\mu}$ disappearance. The results are consistent with no short baseline disappearance of $\bar\nu_{\mu}$ and we have dramatically improved on the excluded regions of the oscillation parameter space.  MiniBooNE and SciBooNE have pushed the limit on short baseline disappearance of $\bar\nu_{\mu}$ down to roughly $10\%$, the region of interest for sterile neutrino models.

\begin{acknowledgments}
We wish to acknowledge the support of Fermilab, the U.S. Department of Energy, and the National Science Foundation in the construction, operation, and data analysis for the MiniBooNE and SciBooNE experiments. The SciBooNE detector was mainly constructed and operated by the budget of Japan-U.S. Cooperative Science Program. We acknowledge the support of MEXT and JSPS (Japan) with the Japan/U.S. Cooperation Program. We also acknowledge the Los Alamos National Laboratory for LDRD funding. We acknowledge the Physics Department at Chonnam National University, Dongshin University, and Seoul National University for the loan of parts used in SciBar and the help in the assembly of SciBar. We wish to thank the Physics Departments at the University of Rochester and Kansas State University for the loan of Hamamatsu PMTs used in the MRD. We gratefully acknowledge the support of grants and contracts from the INFN (Italy), the Ministry of Science and Innovation and CSIC (Spain), and the STFC (UK). We acknowledge the support by MEXT and JSPS with the Grant-in-Aid for Scientific Research A 19204026, Young Scientists S 20674004, Young Scientists B 18740145, Scientific Research on Priority Areas ``New Developments of Flavor Physics," the global COE program ''The Next Generation of Physics, Spun from Universality and Emergence," and the Japan-U.S. Cooperative Science Program between JSPS and NSF.
\end{acknowledgments}

\FloatBarrier

\end{document}